\PassOptionsToPackage{square,numbers}{natbib}
\documentclass[preprint,3p]{elsarticle}
\usepackage{bbm}
\usepackage{booktabs}
\usepackage{multirow}
\usepackage{float}
\usepackage{tikz}
\usetikzlibrary{arrows.meta,
                decorations.pathreplacing,
                    calligraphy,
                positioning}
\usepackage{mathtools}
\usepackage{amsmath}
\usepackage{bm}
\usepackage{amsfonts,dsfont}
\usepackage{indentfirst}
\usepackage{caption}
\usepackage{subcaption}
\usepackage{graphicx}
\usepackage{xcolor}
\usepackage{todonotes}
\usepackage{natbib}
\usepackage{url} 
\usepackage{copyrightbox}
\usepackage{hyperref} 
\usepackage{graphicx}
\usepackage{adjustbox}
\usepackage{amsthm}
\usepackage{circuitikz}
\usepackage{textcomp}
\usepackage[T1]{fontenc}

\DeclarePairedDelimiter\floor{\lfloor} {\rfloor}

\newcommand{\norm}[1]{\left\lVert#1\right\rVert}

\theoremstyle{definition}

\makeatletter
\def\ps@pprintTitle{%
  \let\@oddhead\@empty
  \let\@evenhead\@empty
  \def\@oddfoot{\reset@font\hfil\thepage\hfil}
  \let\@evenfoot\@oddfoot
}
\makeatother

\begin{document}

\begin{frontmatter}



\title{Corrected Support Vector Regression for intraday point forecasting of prices in the continuous power market
}

 \author[PWr]{Andrzej Pu{\'c}} \ead{andrzej.puc@pwr.edu.pl}
 \author[PWr]{Joanna Janczura}\ead{joanna.janczura@pwr.edu.pl}
 \affiliation[PWr]{organization={Wroc{\l}aw University of Science and Technology, Faculty of Pure and Applied Mathematics},
            addressline={Wyb. Wyspia{\'n}skiego 27},
             city={Wroc{\l}aw},
             postcode={50-370},
             country={Poland}}


\begin{abstract}
In this paper, we develop a new approach to the very short-term point forecasting of electricity prices in the continuous market. It is based on the Support Vector Regression with a kernel correction built on additional forecast of dependent variable. We test the proposed approach on a dataset from the German intraday continuous market and compare its forecast accuracy with several benchmarks: classic SVR, the LASSO model, Random Forest and the na\"{i}ve forecast. The analysis is performed for different forecasting horizons, deliveries, and lead times. We train the models on three expert sets of explanatory variables and apply the forecast averaging schemes. 
Overall, the proposed cSVR approach with the averaging scheme yields the highest forecast accuracy, being at the same time the fastest from the considered benchmarks. The highest improvement in forecast accuracy is obtained for deliveries in the morning and evening peaks.
\end{abstract}




\end{frontmatter}

\section{Introduction}
In recent years, we observed a rapid transformation of the overall profile of electricity production in European electricity markets with a growing share of renewable energy sources (RES). This not only makes electricity demand but also supply highly weather dependent and makes prices highly volatile. The effects of RES on energy prices have been widely reported in the literature; see, e.g., \cite{AM, AG, Maciejowska_EE, Kulakov_Ziel}. The growing uncertainty of electricity supply and limited storage possibilities have led to the increasing importance of short-term balancing of positions in electricity markets. The construction of wholesale electricity markets can vary between countries, but generally the main share of electricity traded on the spot markets is sold in the day-ahead market. Typical contracts for the delivery of electricity within a given time period are settled in an auction scheme one day before the delivery. However, since both the future demand and the future supply cannot be exactly determined on the preceding day, the main power exchanges are usually complemented by intraday markets. Since RES generators are often willing to sell energy in short time before delivery and use the opportunity to balance their positions in response to new information, like e.g. more accurate weather forecast, the volume of energy traded in the continuous intraday electricity markets significantly increases in recent years.


The aim of this study is to develop a~method for a~very short-term (intraday) point forecasting of quarter-hourly continuous electricity market prices. Typically, on such a~market, the participants can submit their offers in the afternoon of the day prior to the delivery and up to several minutes before it. The transaction is made each time two offers meet. This yields a~time series of prices for a given delivery, in contrast to auction markets, where only one price is settled. 
As a consequence, the forecasting task is much more complicated. 

The number of papers on continuous power markets forecasting is lower than for the auctions markets. To our best knowledge, \cite{Kath_2018} considered the problem of day-ahead point price forecasting on both auction and continuous quarter-hourly markets for the first time in 2018. Authors found that the quarter-hourly day-ahead prices from EXAA (Energy Exchange Austria) are a~strong predictor for the intraday prices for the same quarter-hourly products. A year later, authors of \cite{NARAJEWSKI2020100107} attempted to forecast the ID3, i.e. the volume-weighted average from three hours before the delivery, for both hourly and quarter-hourly prices on the German continuous market. Using LASSO (Least Absolute Shrinkage and Selection Operator) and Elastic Net techniques they performed the forecasts during the day of the delivery, 3 hours 15 minutes before it. They found that the last known continuous intraday market price for the forecasted delivery is the strongest benchmark for hourly ID3  products. None of the considered models managed to outperform this benchmark in the forecast accuracy. However, most of the models outperformed the last known price in the case of quarter-hourly ID3 products. The authors track down this difference in the results to a~low liquidity at the time of forecasting in comparison to the hourly market. Nevertheless, their variable selection algorithm has shown that the most relevant regressors for intraday quarter-hourly ID3 price forecasting are the most recent intraday auction price and the last known price. Similar results were reported on hourly ID3 prices by \cite{UNIEJEWSKI2019} in the same year. Also in the same year, \cite{beating_the_naive} investigated the possible extensions to the hourly ID3 products forecasting by averaging the last known price with the LASSO forecast and augmenting the LASSO model with timely predictions of fundamental variables for the coming hours. Those approaches allowed to outperform the last known price benchmark. Later papers focused mainly on the path and probabilistic aspects of forecasting. Authors of \cite{NARAJEWSKI2020115801} considered the usage of generalized additive model for forecasting of the trajectory of 5-minutely volume-weighted averages from three hours up to 30 minutes before the delivery of hourly intraday products. Although described approach managed to successfully model the volatility in the German continuous intraday market, the authors suggest that using the traded volume and price of the nearby products might further enhance the results. Another insight into the drivers of prices at the continuous market and their volatility was given in \cite{Hirsh_simulation_based}. Authors suggest that the volatility is driven by the merit-order regime and the time to delivery, while the tail of price distribution is mostly influenced by the past price differences and trading activity. On top of that, authors of \cite{hirsch2023multivariatesimulationbasedforecastingintraday} provided an extensive analysis of cross-product price effects and cross-bordershared order books, highlighting the impact of those factors on the volatility of the continuous intraday prices. Finally, \cite{SERAFIN2022106125} proposed a~novel method of probabilistic paths forecasting of the continuous intraday prices using the Gaussian copula to generate paths and adjusted quantile lines to generate the prediction bands. This approach managed to outperform the well-established similar-date benchmark by 25\%. 

In this paper, we focus on the forecasting of minutely volume-weighted price averages of the quarter-hourly electricity prices from the continuous intraday electricity market. The choice of higher granularity leads to higher flexibility, as results can be later aggregated into half-hourly and hourly products. To our best knowledge, all of the previous point forecasting studies in this area focused on ID3 index forecasting or path forecasting at horizons around 3~hours before the delivery. However, with the overall increase in the liquidity of the intraday market, the liquidity of quarter-hourly products increased too. Thus, we argue that it becomes feasible and more important to model and forecast prices in higher granularity on the newer dataset. Furthermore, we propose a~novel, computationally fast solution for very short-term forecasting of electricity prices based on the Support Vector Regression (SVR) \cite{NIPS1996_d3890178} model. The usage of kernel methods in electricity price forecasting literature is limited. Recently, the SVR model was used as a~simpler machine learning comparison to neural networks, see, e.g. \cite{sara_atef__2019} and \cite{NN_vs_SVR}. However, the authors of \cite{TSCHORA2022118752} showed that the multivariate adaptation of SVR, MultiSVR, outperforms both LASSO and dense neural networks in point forecasting of the French day-ahead electricity prices on a~two-year test period from 01.01.2020 to 31.12.2021. In the same paper, another multivariate SVR adaptation, ChainSVR, outperformed the LASSO model by $2.5\%$ in terms of MAE on German day-ahead prices. 
On the other hand, as shown in \cite{beating_the_naive}, the na\"{i}ve forecast, being the last information from the market, is a~very strong predictor in the very short-term forecasting tasks. The authors suggest that to outperform the na\"{i}ve, the LASSO model can be averaged with it. In this paper, we develop this idea further by using the SVR and introducing kernel corrections based on the na\"{i}ve forecast, called later cSVR (corrected SVR). Thus, we show that the results can be further enhanced by using the na\"{i}ve not only as the parameter and one of the averaged forecasts, but also as a~correction to the model architecture. We test our approach on the dataset from the German intraday continuous power market, comparing the cSVR model performance with the last known price, SVR, LASSO and Random Forest benchmarks.

The remainder of the paper is structured as follows. In Section \ref{sec:dataset} we describe the analysed data and the considered market design. We also present the preprocessing techniques and datasets used later to train the proposed models. In Section \ref{sec:models} we introduce all of the considered models for price forecasting and propose a kernel correction approach. Section \ref{sec:forecasting_study} contains a~detailed description of the forecasting study and the obtained results. Finally, Section \ref{sec:summary} concludes the paper.

\section{Dataset}
\label{sec:dataset}
The general design of the German spot electricity market relies mainly on the day-ahead and intraday trading \cite{EPEX_market_design}. The day-ahead market is operated through a~blind auction, where orders are entered by the participants before the order book closure at 12:00 CET. Then, the auction for the day-ahead concludes, and the market clearing price is established, lying at the intersection of supply and demand curves. The delivery periods traded on the day-ahead market are hours and blocks of hours in the next day. However, EXAA (Energy Exchange Austria) handles an earlier auction, where also the quarter-hourly products are traded. Due to the increasing share of renewable energy sources, the balancing of demand and supply also requires the intraday market. There are two types of intraday markets in Germany: auction and continuous. On the auction market, quarter-hourly, hourly and block products are traded for the next day at 15:00 CET. Continuous intraday trading for the same products starts at 16:00 CET and continues up to 5 minutes before delivery \cite{Next-Kraftwerke}. A~crucial element of all these markets is that they allow cross-border trading. The German day-ahead market is integrated into the Single Day-Ahead Coupling (SDAC) and intraday into the Single Intraday Coupling (SIDC) encompassing across 27 and 25 European countries, respectively \cite{SIDC}.

The dataset analysed in this paper consists of quarter-hourly prices from the German continuous intraday market and spans from 1 November 2018 to 31 December 2020. Each transaction is characterised by 
time of trade, price in euros, volume 
and delivery period specification. The ticks of trade time, price, and volume are respectively one second, euro cent, and 0.1MW. 

\begin{figure}[!h]
    \centering
    \includegraphics[width=\textwidth]{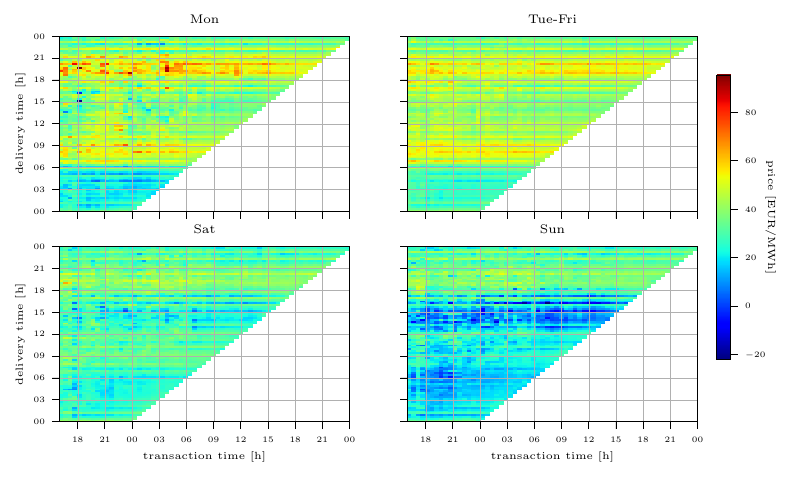}
    \caption{Price averages over the whole dataset and 30 minutes intervals of transaction time. The averages were separately calculated for deliveries in Mondays (top, left panel), Tuesdays-Fridays (top, right panel), Saturdays (bottom, left panel) and Sundays (bottom, right panel) to illustrate weekly patterns.}
    \label{fig:avg_3D}
\end{figure}

\begin{figure}[!h]
    \centering
    \includegraphics[width=\textwidth]{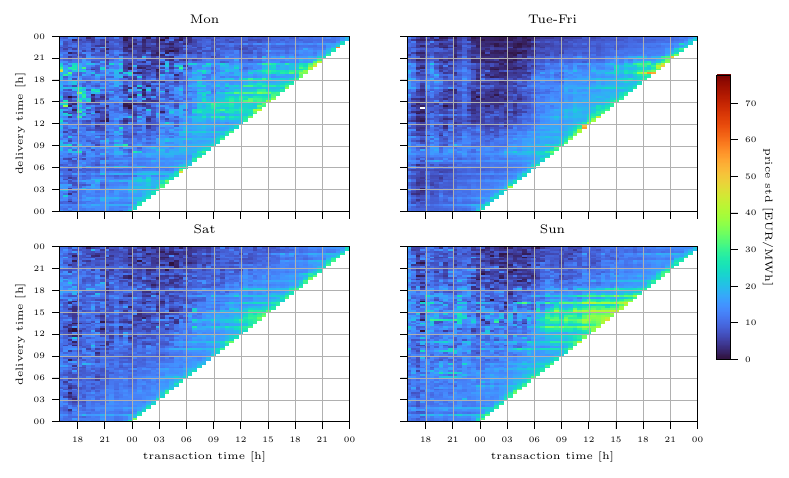}
    \caption{Standard deviation of transaction prices in one minute averaged over 30 minutes intervals of transaction time and the whole dataset. The standard deviations were separately calculated for deliveries in Mondays (top, left panel), Tuesdays-Fridays (top, right panel), Saturdays (bottom, left panel) and Sundays (bottom, right panel) to illustrate weekly patterns.}
    \label{fig:std_3D}
\end{figure}

\begin{figure}[!h]
    \centering
    \includegraphics[width=\textwidth]{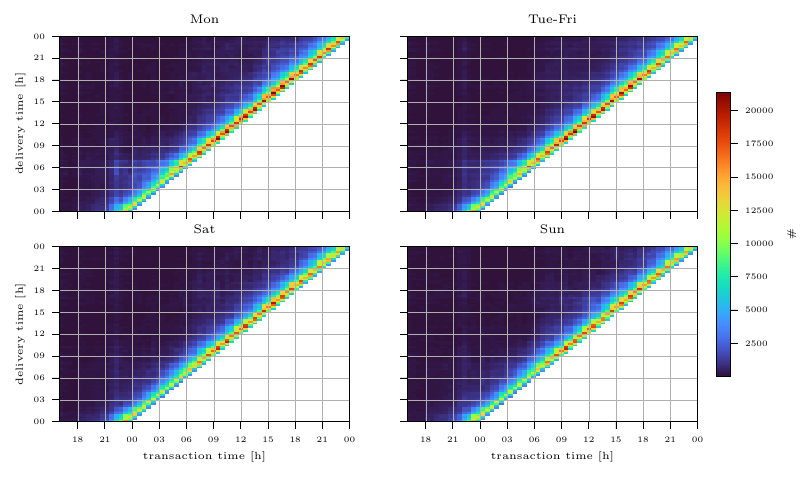}
    \caption{Total number of transactions in every 30 minutes trade time interval. The number of transactions were separately calculated for deliveries in Mondays (top, left panel), Tuesdays-Fridays (top, right panel), Saturdays (bottom, left panel) and Sundays (bottom, right panel) to illustrate weekly patterns.}
    \label{fig:trades_no_3D}
\end{figure}

The general behaviour of the analyzed transaction data is visualised in Figures \ref{fig:avg_3D}-\ref{fig:trades_no_3D} in terms of the average price, its standard deviation, and the number of trades recorded within 30 minutes intervals. Looking at the average prices for all 96 quarter-hourly deliveries (Figure \ref{fig:avg_3D}), clear differences between peak and off-peak hours can be observed. Products in peak hours (8-20) are priced on average higher than products in the night and early morning. The morning and evening peak periods are also priced higher than the rest of the peak. In the morning, the price is on average higher between 7:00 and 12:00. In the evening, we observe a~higher price between 18:00 and 21:00. This is the same residual load driven behaviour as in the case of the German day-ahead and auction intraday prices, see e.g. \cite{janczura_puc_strategie}. 
The standard deviation of the transaction prices, averaged across 30 minutes transaction time intervals, is presented in Figure \ref{fig:std_3D}. During the working days, the peak hours are more volatile and the volatility increases closer to delivery. Less volatile transaction prices are observed during the night on the day of delivery. Those periods occur for both working days and weekends for transaction times after midnight and less than approximately 3 hours before the delivery. 
The number of trades is presented in Figure \ref{fig:trades_no_3D}. We observe the highest number of trades between 1 hour and 30 minutes before delivery. Overall, the density of trades with respect to trade time increases close, i.e. around 3 hours, to delivery. A slight increase in volumes is visible earlier, mostly for morning and evening peaks. Moreover, a~higher trades density is observed around 10 p.m. on the day prior to delivery. 

Analysis of mean, standard deviation, and number of trades shows that the period from 3 hours to 30 minutes before delivery is characterised by the highest liquidity. Simultaneously, the scarcity of trades performed more than 3 hours before delivery suggests that the task of accurate point forecasting on such horizons may be infeasible. We will use these observations in the simulation study construction.

\begin{figure}[!h]
    \centering
    \includegraphics[width=\textwidth]{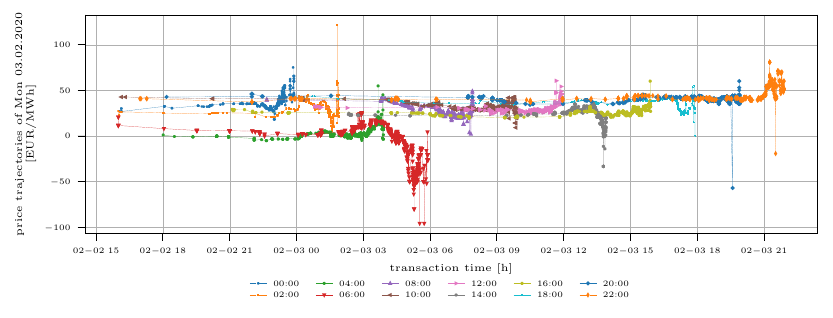}
    \caption{Prices trajectories for chosen Monday, 03.02.2020, deliveries.}
    \label{fig:example_of_trajectories}
\end{figure}

To better understand the time evolution of intraday market prices, beyond the further aggregated data, in Figure \ref{fig:example_of_trajectories} we show an example of the trajectories of the chosen deliveries on Monday, 03.02.2020. The behaviour of trajectories supports previous conclusions: majority of the trades occur closer than 3 hours before the delivery. Moreover, the increase in volatility is visible closer to the traded product. Note also the frequent occurrence of prices at a~very different level. One may consider those as outliers. However, the fact that they occur in a~high proportion of the trajectories suggests that these trades may also carry a~significant information about the market. Another important remark, and at the same time a~feature of continuous market, is that the first transaction for a given delivery is not necessarily settled at 16:00, i.e. at the trade starting time.

For the construction of price forecasts, we also use exogenous variables. This is a common approach in the electricity price forecasting literature, as prices are strongly dependent on the general physical characteristics of the market, see \cite{hong_weron_review_2020, GIANFREDA20122228}, and the weather conditions that a~given market is facing, see \cite{Weron_review}. The considered exogenous variables match the time range of the transaction dataset and include cross-border physical flow between France and Germany, RES generation, day-ahead RES generation forecasts, total load, day-ahead total load forecasts, day-ahead electricity prices, and intraday auction electricity prices. The list of exogenous variables with their units, frequency, and source is given in Table \ref{tab:ex_var}. The choice of France connection in cross-border flows was dictated by the proportion of DE--FR exchange in total imports: $29.7$\% in 2018, $29.6$\% in 2019, and $22.5$\% in 2020 \cite{germany_exchange}. In the considered dataset, the exchange with France is responsible for 2.7\% of imported and 3.3\% of exported volume. Note that further we distinguish between import and export by assigning positive values of volume to the energy exported from Germany and negative to the imported volume.
\begin{table}[!h]
\caption{A list of exogenous variables used for forecast construction together with their units, frequencies and sources. }
\label{tab:ex_var}
\begin{adjustbox}{max width=\textwidth}
\begin{tabular}{|l|l|l|l|l|}
\hline
Data                                         & Unit      & Notation & Frequency     & Source                                                                                                                                                                                                                                                       \\ \hline
Cross-border physical flow between FR and DE & [MW]      & $X^1$    & hourly        & \href{https://transparency.entsoe.eu/transmission-domain/physicalFlow/show?name=&defaultValue=false&viewType=TABLE&areaType=BORDER_BZN&atch=false&dateTime.dateTime=01.01.2020+00:00|CET|DAY&border.values=CTY|10Y1001A1001A83F!BZN_BZN|10Y1001A1001A82H_BZN_BZN|10YFR-RTE------C&dateTime.timezone=CET_CEST&dateTime.timezone_input=CET+(UTC+1)+/+CEST+(UTC+2)}{ENTSO-E} \cite{ENTSOE}                                                                                                                                                                                                   \\ \hline
RES generation - actual                      & [MW]      & $X^2$    & quarter-hourly & \href{https://transparency.entsoe.eu/generation/r2/actualGenerationPerProductionType/show?name=&defaultValue=true&viewType=TABLE&areaType=BZN&atch=false&datepicker-day-offset-select-dv-date-from_input=D&dateTime.dateTime=01.01.2020+00:00|CET|DAYTIMERANGE&dateTime.endDateTime=01.01.2020+00:00|CET|DAYTIMERANGE&area.values=CTY|10Y1001A1001A83F!BZN|10Y1001A1001A82H&productionType.values=B01&productionType.values=B02&productionType.values=B03&productionType.values=B04&productionType.values=B05&productionType.values=B06&productionType.values=B07&productionType.values=B08&productionType.values=B09&productionType.values=B10&productionType.values=B11&productionType.values=B12&productionType.values=B13&productionType.values=B14&productionType.values=B20&productionType.values=B15&productionType.values=B16&productionType.values=B17&productionType.values=B18&productionType.values=B19&dateTime.timezone=CET_CEST&dateTime.timezone_input=CET+(UTC+1)+/+CEST+(UTC+2)}{ENTSO-E} \cite{ENTSOE} \\ \hline
RES generation - day-ahead                   & [MW]      & $X^3$    & quarter-hourly & \href{https://transparency.entsoe.eu/generation/r2/dayAheadGenerationForecastWindAndSolar/show?name=&defaultValue=true&viewType=TABLE&areaType=BZN&atch=false&dateTime.dateTime=01.01.2020+00:00|CET|DAYTIMERANGE&dateTime.endDateTime=01.01.2020+00:00|CET|DAYTIMERANGE&area.values=CTY|10Y1001A1001A83F!BZN|10Y1001A1001A82H&productionType.values=B16&productionType.values=B18&productionType.values=B19&processType.values=A18&processType.values=A01&processType.values=A40&dateTime.timezone=CET_CEST&dateTime.timezone_input=CET+(UTC+1)+/+CEST+(UTC+2)}{ENTSO-E} \cite{ENTSOE}                                                                                           \\ \hline
Total load - actual                          & [MW]      & $X^4$    & quarter-hourly & \href{https://transparency.entsoe.eu/load-domain/r2/totalLoadR2/show?name=&defaultValue=false&viewType=TABLE&areaType=BZN&atch=false&dateTime.dateTime=01.01.2020+00:00|CET|DAY&biddingZone.values=CTY|10Y1001A1001A83F!BZN|10Y1001A1001A82H&dateTime.timezone=CET_CEST&dateTime.timezone_input=CET+(UTC+1)+/+CEST+(UTC+2)}{ENTSO-E} \cite{ENTSOE}                                                                                                                                                                                                                                                                                                                                                                                                                                                                                                                               \\ \hline
Total load - day ahead                       & [MW]      & $X^5$    & quarter-hourly & \href{https://transparency.entsoe.eu/load-domain/r2/totalLoadR2/show?name=&defaultValue=false&viewType=TABLE&areaType=BZN&atch=false&dateTime.dateTime=01.01.2020+00:00|CET|DAY&biddingZone.values=CTY|10Y1001A1001A83F!BZN|10Y1001A1001A82H&dateTime.timezone=CET_CEST&dateTime.timezone_input=CET+(UTC+1)+/+CEST+(UTC+2)}{ENTSO-E} \cite{ENTSOE}                                                                                                                                                                                                                                                                                                                                                                                              \\ \hline
Day-ahead prices                             & [EUR/MWh] & $X^6$    & quarter-hourly & \href{https://transparency.entsoe.eu/transmission-domain/r2/dayAheadPrices/show?name=&defaultValue=true&viewType=GRAPH&areaType=BZN&atch=false&dateTime.dateTime=01.01.2020+00:00|CET|DAY&biddingZone.values=CTY|10Y1001A1001A83F!BZN|10Y1001A1001A82H&resolution.values=PT15M&resolution.values=PT30M&resolution.values=PT60M&dateTime.timezone=CET_CEST&dateTime.timezone_input=CET+(UTC+1)+/+CEST+(UTC+2)}{ENTSO-E} \cite{ENTSOE}                                                                                                                                                                         \\ \hline
Intraday auction prices                             & [EUR/MWh] & $X^7$    & quarter-hourly & EPEX \cite{EPEX}                                                                                                                                                                                                                                                                                                         \\ \hline
\end{tabular}%
\end{adjustbox}
\end{table}

\begin{figure}[!h]
    \centering
    \includegraphics[width=\textwidth]{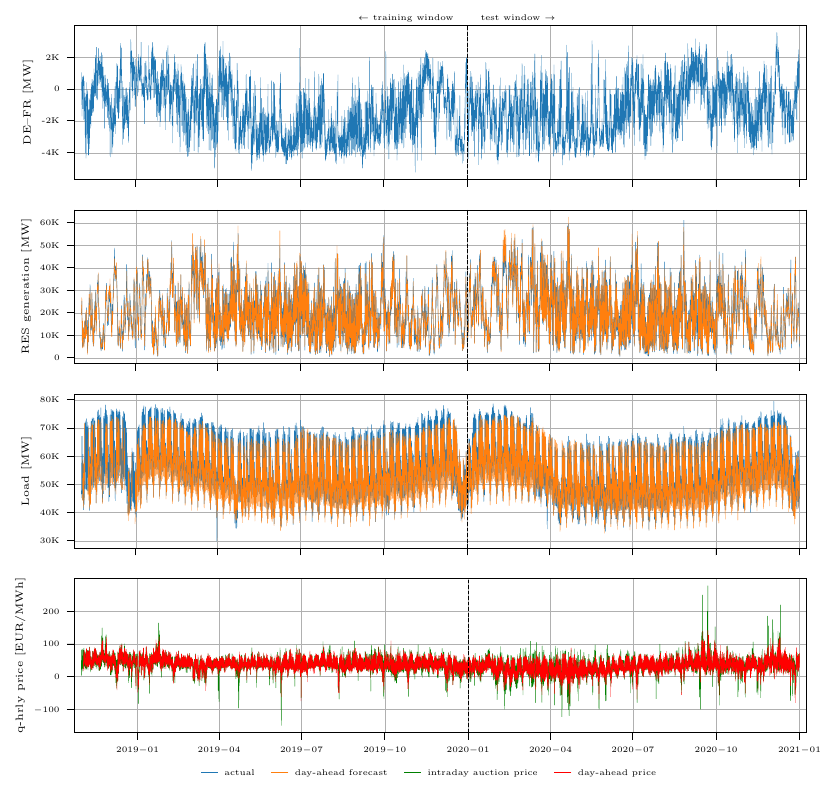}
    \caption{Exogenous variables, i.e. cross-border physical flow between France and Germany (top panel); RES generation and day-ahead RES generation forecasts; total load and day-ahead total load forecasts; day-ahead electricity prices and intraday auction electricity prices in the considered period.}
    \label{fig:exog}
\end{figure}

The considered exogenous variables are plotted in Figure \ref{fig:exog}. 
Note that the majority of flows are imports from France. In fact, the proportion of FR \textrightarrow DE to DE \textrightarrow FR total exchanged power over the considered period is 3 to 1. The RES generation volume slightly increases over time, which is coherent with the increasing installed capacity of the RES. Also, a drop of load values is visible in March and April 2020. This was caused by the initial COVID-19 lock-downs. The seasonal load drop in December is caused by the holidays. Looking at the quarter-hourly day-ahead prices, the quantity of negative spikes is also larger during the first months of the COVID-19 pandemic, and the volatility of prices increases afterwards.

Obviously, when constructing forecasts, only the data available at the time of forecasting can be used. According to ENTSO-E \cite{physical_flows_data_avails}, all considered exogenous data is available no later than one hour after the end of the application period. The delay in continuous intraday data availability on EPEX SPOT sFTP server is 20 minutes, {see \cite{epex_transactions_delay}}. Hence, we set the delay for intraday exogenous information used in the model and prices accordingly. 
Note that there is no need to apply any delay to quarter-hourly day-ahead prices as those are settled at EXAA (Energy Exchange Austria) during the 10:15 a.m. auction one day prior to the delivery. Similarly, the auction intraday prices are set at 15:00. on the day prior to the delivery at EPEX. 

\begin{figure}[!h]
    \centering
    \includegraphics[width=\textwidth]{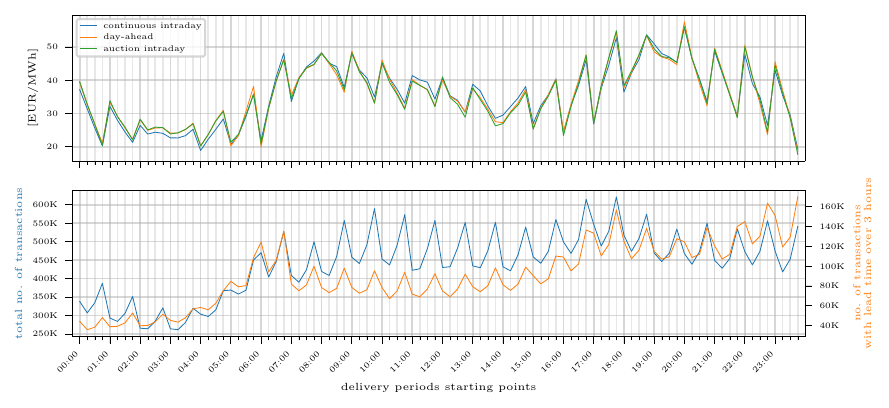}
    \caption{Top panel: auction intraday, continuous intraday  and day-ahead averaged electricity prices for all 96 quarter-hourly products. Bottom panel: total transaction number (left axis) and transaction number limited to lead time being longer than three hours (right axis). Note that the curves in the bottom panel are plotted in different scales in order to clearly reflect their shape.}
    \label{fig:intraday_vs_da_qtrhrly}
\end{figure}

The choice of EXAA quarter-hourly day-ahead prices as a~regressor instead of a~more recent information from the hourly auction at 12:00 is dictated by a~visible difference in price levels and liquidity between quarters, see Figure \ref{fig:intraday_vs_da_qtrhrly}. In the upper panel, we compare the average prices from three quarter-hourly markets. Day-ahead and auction intraday markets prices are averaged across days in the sample, and the continuous intraday price is averaged across days and transactions. The price evolution throughout the day and the scheme within hours are similar in all three markets. Interestingly, the first quarter is priced higher than the rest from 8:00 to 13:00 and from 20:00 to 2:00. For the other hours, the highest energy price is settled in the last, fourth quarter. The authors of \cite{KIESEL201777} provide a~wider explanation of this behaviour during the day based on the dependencies between quarter-hourly and hourly products. For hours between 8:00 and 14:00, the RES providers sell energy for the full hour at the day-ahead market, and as the sun is going up, they are not able to provide the energy for all quarters. Therefore, they balance their hourly position by buying in the first quarter hour, which moves its price up. But, as the sun is, on average, going up, they can also produce too much in the last quarter hour. This moves the price down as more providers want to sell energy. After 14:00 the sun is going down, so the process reverses, and now the price from the first quarter is the lowest, and the price from the last quarter is the highest. The authors of \cite{KIESEL201777} also provide a~similar explanation for a zigzag pattern in night hours, this time driven not by RES production but by the relations between fossil power plants' production and demand. To sum up, the zigzag pattern of quarter-hourly prices is caused by the balancing of positions within hourly products. To further investigate this behaviour, in the bottom panel of Figure \ref{fig:intraday_vs_da_qtrhrly} we compare the total number of transactions with the number of transactions settled earlier than three hours before delivery. 
Note that almost always the liquidity is highest during the last quarter of an hour. This suggests that balancing hourly products in the last quarter is very important for market participants. Another important remark is that trading starts earlier 
- for deliveries between the first off-peak and the morning peak, i.e. from 5:45 to 6:45 a.m. and later before the evening peak, i.e. from 16:45 to 18:00. A~high number of early transactions are also observed for deliveries from 22:45 to midnight, which could be associated with balancing of daily contracts. Overall, this earlier start of trading is observed for parts of the day with the highest intra-hourly variability in the first panel of Figure \ref{fig:intraday_vs_da_qtrhrly}.

\subsection{Data preprocessing}
\label{preprocessing}

\begin{figure}[!h]
    \centering
    \includegraphics[width=\textwidth]{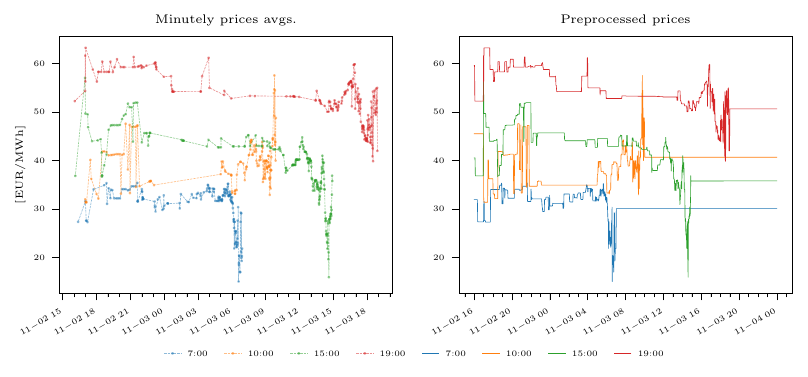}
    \caption{Sample transaction prices for quarter-hourly deliveries on 11-03-2019 starting at 7:00, 10:00, 15:00, 19:00 (left panel) and the corresponding preprocessed price trajectories (right panel).}
    \label{fig:price_preprocessed_trajectories}
\end{figure}

Transaction data for continuous market is recorded for each second in the trading period. However, in this study, we aggregate prices for a~given product into minutely bins. This allows us to avoid including a large number of time points with no transactions, but at the same time, minutely data still reflects high-frequency dynamics of prices. 
Precisely, for each delivery $d$ we calculate a~volume-weighted average of transaction prices from each minute. If there are no trades at the beginning of the starting period (16:00-16:01), we put the last available market information about the price for delivery $d$, i.e. the auction intraday price settled at 15:00. 
This is done for each minute before the first transaction for delivery $d$ occurs. Further, to achieve equal lengths of price trajectories for all deliveries and keep possibly meaningful information, we technically put the average price from the last 3 hours prior to the delivery in all minutes in the given day after the delivery. Price trajectories after preprocessing are presented in Figure \ref{fig:price_preprocessed_trajectories}.

\begin{figure}[!h]
    \centering
    \includegraphics[width=\textwidth]{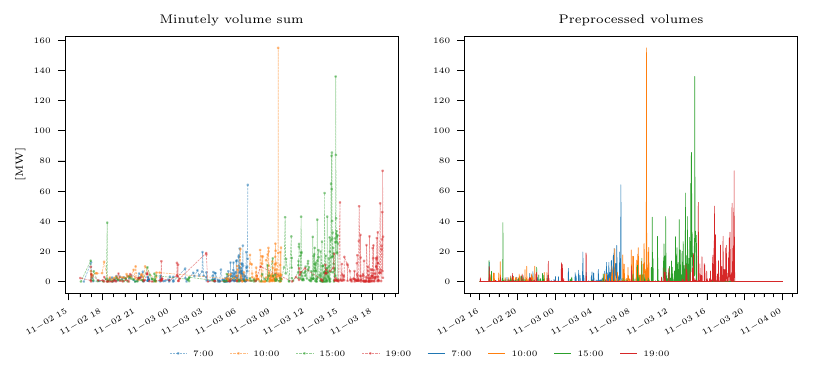}
    \caption{Sample transaction volumes for quarter-hourly deliveries on 11-03-2019 starting at 7:00, 10:00, 15:00, 19:00 (left panel) and the corresponding preprocessed volume trajectories (right panel).}
    \label{fig:volume_preprocessed_trajectories}
\end{figure}

The volume trajectories are preprocessed in a similar way, namely we calculate the average minutely volume. Note that if there were no transactions for delivery $d$ in a given minute, we simply put the value of 0 MWh. Thus, this variable serves a double role: it indicates minutes in which transactions were settled and their total volume. The preprocessed volume trajectories are presented in Figure \ref{fig:volume_preprocessed_trajectories}.


\begin{figure}[!h]
\centering

\begin{tikzpicture}[
    BC/.style = {
        decorate,
        decoration={calligraphic brace, #1,
                    raise=3pt, amplitude=6pt},
                    very thick, thick, 
                    pen colour={black}
                },
    lbl/.style={inner xsep=0pt, text height=1.5ex, text depth=.25ex}    
                        ]
    \draw[-Straight Barb]   (0,0) -- (8,0);
    \foreach \x/\i/\j/\k in {1/m - 20min/, 3/m/,6/s/, 7/d/}
    {
        \draw (\x,3pt)    node [lbl, above] {$\j$}
            -- ++ 
        (0,-6pt)    node (n\x) [lbl, below] {$\i$};
    }
    \draw[BC=mirror]       
            (n1.south) -- node[below=9pt] {data avail. delay} (n3.south);
    
    \draw[BC=mirror]       
            (n6.south) -- node[below=9pt] {lead time} (n7.south);
    
    \draw[-{Stealth[length=3mm]}, bend left=90] (3,0.3) to node[above, sloped] {forecast} (6,0.3);
\end{tikzpicture}

\caption{Forecasting task timeline. Delivery quarter hour is denoted as $d$, transaction time as $s$ and the moment of forecast calculation as $m$. The distance between transaction time and start of delivery is further denoted as lead time.}
\label{fig:forecasting_timeline}
\end{figure}
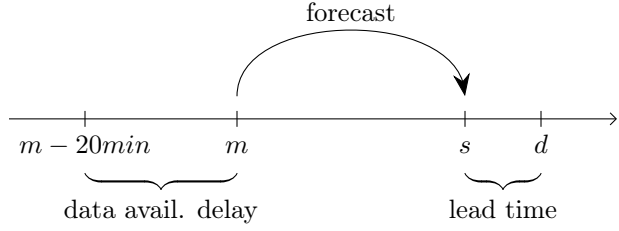

Denote the preprocessed minutely price as $P_{d,T}(s)$ and the corresponding minutely volume as $V_{d,T}(s)$, where $s$~is the transaction time in minutes, $d$~is the delivery in quarter hours and $T$~is the day of delivery. A time distance between trade $s$ and delivery $d$, $d-s$, will be further called the lead time. Our goal is to calculate at a chosen moment $m$, $m < s$, the forecast of the price at time $s$, $P_{d,T}(s)$. Graphical representation of this timeline is presented in Figure \ref{fig:forecasting_timeline}.

To simplify the task, we difference the prices with a~lag of $\delta_{s,m} = s - (m - 20)$ minutes, i.e. the forecasting horizon including a~delay of intraday continuous prices publication. 
The resulting differenced price and volume are given by

\begin{equation}
\begin{split}
    &P'_{d, T}(u) = P_{d, T}(u) - P_{d, T}(u - \delta_{s,m}),\\&
    V'_{d, T}(u) = V_{d, T}(u) - V_{d, T}(u - \delta_{s,m}),
\end{split}
\end{equation}
where $u$ is a minute in a~given trajectory. This way, the differenced variables $P'_{d, T}(u)$ and $V'_{d, T}(u)$, describe the change in price and volume with respect to time $m$. These variables will serve as the basis for the forecasting model construction.  

\subsection{Expert sets of explanatory variables}
\label{sec:expert_sets}

The preprocessed price and volume trajectories consist of 1905 minutely time points for each of 96 quarter-hourly deliveries at each trading day. This multitude of variables requires some initial choice of the data used for the training of the forecasting model. In the following, we suggest three sets of explanatory variables known at time $m$ for $P'_{d,T}(s)$ and give intuition on each choice.

The construction of the first set of variables, denoted by $S_{d,T,m,s}^1$, is based on the assumption that the information needed to forecast the price of delivery $d$ is contained in the trajectories for this and adjacent deliveries. Therefore, we use price trajectories for deliveries from $d-8$ up to $d+4$. We do not consider deliveries from adjacent days, so in the case of the first and last deliveries of the day, this range is cut accordingly. 
Note that we can only use the information available up to the moment of forecast calculation $m$. Therefore, each of those trajectories spans from 16:00 on the previous day up to the moment $m - 20$. In addition, to limit the number of variables in this set, we sample each trajectory in quarter-hourly ($15$ minutes) intervals. If the length of the trajectory is not divisible by $15$, we add the most recent, so possibly the most important, value to the resulting resampled trajectory. Let us denote those resampled time steps as a~vector~$\hat{\bm{u}}$.

\begin{equation}
    \hat{\bm{u}} = \begin{cases}
			(0, 15, \dots, \floor*{\frac{m - 20}{15}}\cdot 15, m - 20), & \text{if $(m - 20) \bmod 15 > 0$},\\
            (0, 15, \dots, m - 20), & \text{otherwise},
		 \end{cases}
\end{equation}
where $\floor*{x} $ denotes the highest integer value lower or equal $x$ and $x$ mod $y$ is the remainder of the division $\frac{x}{y}$. After resampling, we drop all trajectories with zero variance. We perform this step to eliminate unnecessary information coming from trajectories with no trades. Recall that such constant trajectories are technically filled with quarter-hourly auction intraday price (see Section \ref{preprocessing} for details). Further, to account for the recent market behaviour at the considered trade time $s$ and delivery $d$, we include  also the corresponding continuous intraday prices from the last seven days, i.e. $P'_{d,T-7}(s)$, $P'_{d,T-6}(s)$, $\dots$, $P'_{d,T-1}(s)$.  
From the volume trajectories we extract the last known value of the change of the total traded volume, $V'_{T}(m - 20) = \sum_{d=1}^{96} V'_{d,T}(m - 20)$. This value describes the change in the amount of power that was traded in minute {$m-20$} for all deliveries in day $T$. To supplement it, we also include the sum of the volumes traded in the recent 60 minutes for the delivery $d$, $\sum_{u = m-80}^{m-20} V_{d, T}(u)$ and the total number of minutes with any transactions in the recent hour $N_{d, T}(m-20) = \sum_{u = m-80}^{m-20} \mathbbm{1}_{V_{d, T}(u) > 0}$. These variables serve as a~proxy for the  market liquidity.

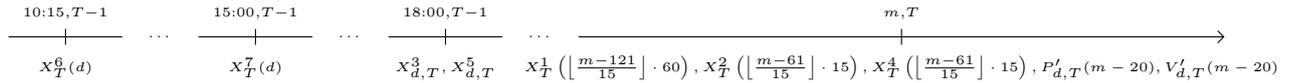
\begin{figure}[H]
\centering
\begin{tikzpicture}[
    BC/.style = {
        decorate,
        decoration={calligraphic brace, #1,
                    raise=3pt, amplitude=6pt},
                    very thick, thick, 
                    pen colour={black}
                },
    lbl/.style={inner xsep=0pt, text height=1.5ex, text depth=.25ex}    
                        ]
    \tikzstyle{every node}=[font=\tiny]
    \draw[-]   (0,0) -- (1.5,0);

    \node[rectangle, draw=none, minimum size=1pt] at (2, 0) {\dots};

    \draw[-]   (2.5,0) -- (4,0);

    \node[rectangle, draw=none, minimum size=1pt] at (4.5, 0) {\dots};

    \draw[-]   (5,0) -- (6.5,0);

    \node[rectangle, draw=none, minimum size=1pt] at (7, 0) {\dots};

    \draw[-Straight Barb]   (7.5,0) -- (16,0);

    \foreach \x/\i/\j/\k in {0.75/X_{T}^6(d)/{\scriptstyle\text{10:15}, T - 1}, 3.25/X_{T}^7(d)/{\scriptstyle\text{15:00}, T - 1}, 5.75/{X_{d,T}^3, X_{d,T}^5}/{\scriptstyle\text{18:00}, T-1}, 11.75/{X_{T}^1\left(\left\lfloor \frac{m-121}{15} \right\rfloor \cdot 60\right), X_{T}^2\left(\left\lfloor \frac{m - 61}{15} \right\rfloor \cdot 15\right), X_{T}^4\left(\left\lfloor \frac{m - 61}{15} \right\rfloor \cdot 15\right), P'_{d,T}(m-20), V'_{d,T}(m-20)}/{\scriptstyle m, T}}
    {
        \draw (\x,3pt)    node [lbl, above] {$\j$}
            -- ++ 
        (0,-6pt)    node (n\x) [lbl, below] {$\i$};
    }

\end{tikzpicture}
\caption{Timeline of data availability for all exogenous variables used in the study. Time moments indicating when the data is available for forecasting at time $m$ are provided above the axis.}
\label{fig:exog_variab_timeline}
\end{figure}

In addition to the transaction data variables, we also include system-related exogenous variables (see Table \ref{tab:ex_var}) to supplement the information contained in recent trades with a~broader view of the market. Namely, taking into account delays in data availability, we create 
a~vector of exogenous variables $\mathbf{X}_{d, T, m}$  containing: 
\begin{enumerate}
    \item the last known actual value of cross border physical flow between FR and DE, $X^1_{T}\left(\left\lfloor \frac{m - 121}{60} \right\rfloor \cdot 60\right)$;
    \item the last known actual value of RES generation $X^2_{T}\left(\left\lfloor \frac{m - 61}{15} \right\rfloor \cdot 15\right)$ and its day-ahead forecast for the moment of delivery~$X^3_{d,T}$;
    \item the last known actual value of load $X^4_{T}\left(\left\lfloor \frac{m - 61}{15} \right\rfloor \cdot 15\right)$ and its day-ahead forecast for the moment of delivery~$X^5_{d,T}$;
    \item the last known error of the RES generation forecast, i.e. $X^2_{T}\left(\left\lfloor \frac{m - 61}{15} \right\rfloor \cdot 15\right) - X^3_{\left\lfloor \frac{m - 61}{15} \right\rfloor \cdot 15, T}$ and the last known error of the load forecast $X^4_{T}\left(\left\lfloor \frac{m - 61}{15} \right\rfloor \cdot 15\right) - X_{\left\lfloor \frac{m - 61}{15} \right\rfloor \cdot 15, T}^5$;
    \item day-ahead quarter-hourly prices for the moment of delivery $X_{d,T}^6$;
    \item auction intraday quarter-hourly prices for the moment of delivery $X_{d,T}^7$.
\end{enumerate}
The assumed delay for ENTSO-E data availability is set to 60 minutes, which is the maximum delay value according to the ENTSO-E documentation \cite{ENTSOE}. Thus, for data with hourly granularity, we assume that the most recent known actual value was published at $\left\lfloor \frac{m - 121}{60} \right\rfloor \cdot 60$, while for the quarter-hourly granularity at $\left\lfloor \frac{m - 61}{15} \right\rfloor \cdot 15$. We acknowledge the limitation of those assumptions. In practice, the model should be set up with a~listener that would pick up the latest information and use it in training and forecasting accordingly. Note that the delay in cross-border physical flow data, $m-121$, accounts for the hourly ENTSO-E publication lag, as well as the hourly data granularity. On the other hand, the day-ahead price $X_{T}^6$ is settled for all of the quarter-hourly deliveries $d$ at 10:15 and the auction intraday price, $X_{T}^7$, is settled at 15:00 on the day preceding the delivery. The RES generation and load forecasts for all deliveries $d$, $X_{d,T}^3$ and $X_{d,T}^5$, are published at 18:00 on the day prior to the delivery. Therefore, these variables do not require any lagging. For a~complete timeline of the exogenous data availability, see Figure \ref{fig:exog_variab_timeline}.

Since the weekly price pattern is a standard feature of electricity prices (see Figures \ref{fig:avg_3D}-\ref{fig:trades_no_3D} for illustration), we also use a dummy variable indicating the day of the week. We denote this variable as a~vector $\mathbf{D}_T = (\mathbbm{1}_{\{T\text{ is Monday}\}}, \dots, \mathbbm{1}_{\{T\text{ is Sunday}\}})$. Finally, we add the last known undifferentiated price $P_{d,T}(m - 20)$, which completes the construction of $S^1$.  
\begin{equation}
    \begin{split}
        S^1_{d, T, m, s} = \Bigl\{& \left \{P'_{d-8,T}(\hat{u}_i), \hat{u}_i\in\hat{\bm{u}}\right \}, \left \{P'_{d-7,T}(\hat{u}_i), \hat{u}_i\in\hat{\bm{u}}\right \}, \dots, \left \{ P'_{d+4,T}(\hat{u}_i), \hat{u}_i=\in\hat{\bm{u}}\right \}, 
        \\& P'_{d,T-7}(s),P'_{d,T-6}(s), \dots, P'_{d,T-1}(s),\\&P_{d,T}(m - 20),\\&V'_T(m - 20), \sum_{u = m-80}^{m-20} V_{d, T}(u), N_{d, T}(m-20), \\&\mathbf{X}_{d,T,m}, \mathbf{D}_T \Bigr\}.
    \end{split} \label{eq:S1}
\end{equation}

As a more simplistic solution to price forecasting, we propose an approach based only on recent price values. In the second set of variables $S_{d,T,m,s}^2$, we consider delivery $d$ and deliveries from the previous hour, as well as transactions up to 60 minutes prior to the moment of forecasting. Similarly to $S_{d,T,m,s}^1$, we drop the trajectories with zero variance. 
The resulting variables set $S_{d,T,m,s}^2$ is defined as
\begin{equation}
    \begin{split}
    S^2_{d, T, m, s} = \Bigl\{&\left\{P'_{d-4,T}\left (m-20-\tilde{u}\right ), \tilde{u}= 0,1,\dots,60 \right\}, \left\{P'_{d-3,T}\left (m-20-\tilde{u}\right ), \tilde{u}= 0,1,\dots,60 \right\}, \dots, \\
    &\left\{P'_{d,T}\left (m-20-\tilde{u}\right ), \tilde{u}= 0,1,\dots,60\right\}\Bigr\}.
    \end{split}\label{eq:S2}
\end{equation}
Finally, a variables set including only the most recent price and volume supplemented with exogenous variables is proposed. The assumption behind this set construction is that the majority of the information about the market is already contained in the recent transactions. 
Let us denote this set as $S_{d,T,m,s}^3$ and define as
\begin{equation}
    \begin{split}
        S^3_{d, T, m, s} = \Bigl\{&P_{d,T}(m - 20), P'_{d,T}(m - 20),\\&V'_T(m - 20), \sum_{u = m-80}^{m-20} V_{d, T}(u), N_{d, T}(m-20),\\&\mathbf{X}_{d,T,m}, \mathbf{D}_T\Bigr\}.
    \end{split}\label{eq:S3}
\end{equation}

\subsection{Variables standardisation and correlation filter}
The sets of explanatory variables introduced in Section \ref{sec:expert_sets} will be further used for the derivation of price forecasts. Naturally, first, the parameters of a forecasting model need to be calculated in a training window including explanatory variables as well as the corresponding target variable, i.e. the differenced prices. Let us define the training window for forecasting at time $m$ of the price settled at $s$ for delivery in a~quarter hour $d$ and day $T$ as a set of days $\mathcal{T}=\{T-n, T-n+1,...,T-1\}$, where $n$ is the length of the training window. 
In order to alleviate differences in the levels and scale of different variables, we standardise them over days in the training window. First, the target variable, $P^{'}_{d,t}(s), t\in \mathcal{T}$, is standardised as 
\begin{equation}
    \widetilde{P}^{'}_{d,t}(s) = \frac{P^{'}_{d,t}(s) - \mu_{d,s}}{\sigma_{d,s}},
\end{equation}
where $\mu_{d,s}$ is the average of price over days $t \in \mathcal{T}$ and $\sigma_{d,s}$ its standard deviation.
 Next, each of the explanatory variables is standardised in a similar way. Let ${x}^{i, j}_{d,t,m,s}$ be a~single $j$th variable from the set ${S}^i_{d,t,m,s}$ and let $\bm{x}^{i, j}_{d,m,s}=({x}^{i, j}_{d,T-n,m,s}, {x}^{i, j}_{d,T-2,m,s},...,{x}^{i, j}_{d,T,m,s})$ be a vector of its values in the training window. Note that here also the values from  $S^i_{d,T,m,s}$ are included as these are already known before the forecast calculation. Then the standardised explanatory variables are given by
\begin{equation}
    \widetilde{\bm{x}}^{i,j}_{d,m,s} = \frac{\bm{x}^{i, j}_{d,m,s} - \mu^{i,j}_{d,m,s}}{\sigma^{i,j}_{d,m,s}}, \label{eq:standardisation}
\end{equation}
where $\mu^{i, j}_{d,m,s}$ is the average of the elements of $\bm{x}^{i, j}_{d,m,s}$ and $\sigma^{i, j}_{d,m,s}$ the corresponding standard deviation. Finally, denote the $i$th set of the standardised explanatory variables from the training window as $\tilde{\mathcal{S}}^i_{d,m,s}, i=1,2,3$. 

Before training the model parameters, we also limit the duplicated information using a~correlation-based filtering. It is applied to all explanatory variables from the training sets $\widetilde{\mathcal{S}}^i_{d,m,s}$, $i = 1,2$, except for the exogenous variables listed in Table \ref{tab:ex_var}.
 Specifically, we search for variables correlated stronger than a~given threshold and then remove one of the variables in a strong correlation pair. For set $\widetilde{\mathcal{S}}^1_{d,m,s}$ we use an arbitrary threshold of $0.8$ and for set $\widetilde{\mathcal{S}}^2_{d,m,s}$ a~threshold of $0.95$ for the absolute value of the Pearson correlation coefficient. The choice of higher rejection threshold in the case of $\widetilde{\mathcal{S}}^2_{d,m,s}$ set is dictated by the already limited amount of variables in this set. Note that, since the main goal of filtering is to reject highly correlated price information, the set $\widetilde{\mathcal{S}}^3_{d,m,s}$ is not subject to filtering as it already consists of a~hand picked set of exogenous variables and the last known price. 

\begin{figure}[!h]
    \centering
    \includegraphics[width=\textwidth]{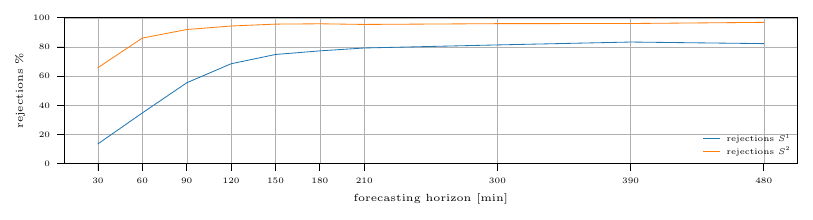}
    \caption{Percentage of variables rejected by the correlation filter in set $S^1$ and $S^2$ for 60 minutes lead time. The forecasting horizon in our setting is shown in Figure \ref{fig:forecasting_timeline}.}
    \label{fig:correlation_filtering_results}
\end{figure}

The results of the filtering for a~lead time of 60 minutes and several forecasting horizons are presented in Figure \ref{fig:correlation_filtering_results}. For shorter forecasting horizons, i.e. 30 to 90 minutes, the rejection ratio increases sharply. For longer horizons, the rejection rates stabilise close to the thresholds, which suggests a~high concentration of correlation values near them. This nondecreasing profile of rejection ratios over forecasting horizons can be explained by market liquidity, which is decreasing with increasing horizon and stagnating at a~very low level over the horizon of 120~minutes.   

\section{Models specification}
\label{sec:models}
The number of individual point forecasts that need to be performed during the forecasting study on continuous markets is incomparably higher than in the case of auction markets analysis. Therefore, the choice of methods must be dictated by a~trade-off between the expected accuracy and computational complexity. Our model of choice is the SVR, which is compared in the paper with the LASSO and the Random Forest (RF). In this section, we provide a~brief theoretical background on each of these models and give a~short intuition on their advantages and disadvantages in our use case study. We also present a~detailed description of how those models were applied in our setting.

\subsection{Support Vector Regression}

The preprocessing technique described in Section \ref{preprocessing} produces time series from data of uneven spacing. The resulting trajectories are very volatile, and therefore our objective is to construct a~model that will be able to capture the nonlinearities of those trajectories with minimum assumptions and computational time required. Models commonly used to tackle such problems are neural networks, which are capable of outperforming the statistical methods in time series forecasting tasks, see e.g. \cite{NN_in_point_time_series, marcjasz_NN_probabilistic}. However, due to the amount of point forecasts that are of interest in the continuous market setting, the computational time required by neural networks makes their usage impractical for very short-term (intraday) forecasting. Thus, we search for methods of less computational complexity. The authors of \cite{jacot_ntk} showed that, in specific cases, neural networks can be approximated by kernel methods based on the neural tangent kernel (NTK). Furthermore, the limit NTKs for dense neural networks were shown to be similar to the Laplace kernel \cite{laplace_NTK_similarity}. Hence, in the paper, we choose a~fast, non-parametric model, specifically the SVR with the Laplace kernel, which belongs to a~wider group of kernel methods. The idea behind kernel methods is to apply a~feature map to transform problem into a~higher dimensional space, where it can be solved by fitting a~linear function. Let us denote this map as $\phi \colon \mathbb{R}^d \to \mathcal{H}$, where $\mathcal{H}$ is a~Hilbert space with the inner product $\langle\cdot,\cdot\rangle_\mathcal{H}$. Based on the Representer Theorem, see \cite{representer_theorem}, and the kernel trick, we only need to know the inner products $\left\langle\phi\left(\bm{x}_i\right), \phi\left(\bm{x}_j\right)\right\rangle_{\mathcal{H}}$ for all $i, j = 1,\dots N$ to perform a~linear regression in the Hilbert space. 

Let us denote a~kernel, i.e. a~symmetric, continuous function, as $K: \mathbb{R}^d \times \mathbb{R}^d \rightarrow \mathbb{R}$ such that $K(\bm{x}_i, \bm{x}_j)=\langle\phi(\bm{x}_i), \phi(\bm{x}_j)\rangle_{\mathcal{H}}$.
We now specify the non-linear SVR model following \cite{NIPS1996_d3890178}. For simplicity of notation, in this paragraph we denote the training vectors as $\bm{x}_i$ and the corresponding target variables as $y_i$, $i = 1, \dots, N$. Our goal is to fit the linear function $f(\phi(\bm{x}_i)) = \bm{w}^T \phi(\bm{x}_i) + b$ in a~Hilbert space resulting from the map $\phi \colon \mathbb{R}^d \to \mathcal{H}$. Here, $\bm{w}$ is the vector of model parameters and $b$ is an intercept. To ensure that the function $f(\cdot)$ is as flat as possible, we search for minimal $\bm{w}^T \bm{w}$. Moreover, we demand that all residues have an absolute value less than $\epsilon$. This creates a~tube of width $2\epsilon$ around the fitted function, in which the penalty resulting from the distance between the points and the function is zero. However, not all points can be contained in this tube, and to deal with infeasible constraints, the model uses slack variables $\bm{\xi}$ and $\bm{\xi}^*$ for the negative and positive residues, respectively. Those are the absolute differences between the training points and the $\epsilon$ tube boundary. Slack variables are zero, if the sample point is within the $\epsilon$ tube. Therefore, the goal of the algorithm is to find a~line in the Hilbert space that contains as many observations as possible inside its $\epsilon$~tube, and the other points are as close to this tube as possible. Furthermore, to control the penalty given to values outside the $\epsilon$ margin, the constant $C$ is introduced. When $C$ is large, the emphasis is placed on the error with respect to the $\epsilon$ tube. In contrast, if $C$ is small, the emphasis is on the weights norm. Following \cite{svr_equation}, the resulting $\epsilon$-Support Vector Regression primal formula is given as

\begin{equation}
    \begin{aligned}
        \min _{\bm{w}, b, \bm{\xi}, \bm{\xi}^*} & \qquad \frac{1}{2} \bm{w}^T \bm{w}+C \left( \sum_{i=1}^N \xi_i+ \sum_{i=1}^N \xi_i^*\right) \\
        \text { subject to } & \qquad \bm{w}^T \phi\left(\bm{x}_i\right)+b-y_i \leq \epsilon+\xi_i, \\
        & y_i-\bm{w}^T \phi\left(\bm{x}_i\right)-b \leq \epsilon+\xi_i^*, \\
        & \xi_i, \xi_i^* \geq 0, i=1, \ldots, N.
    \end{aligned}
    \label{eq:svr_opti}
\end{equation}
Its dual problem is, following \cite{svr_equation}, given by

\begin{equation}
    \begin{aligned}
        \min _{\bm{\alpha}, \bm{\alpha}^*} & \qquad \frac{1}{2}\left(\bm{\alpha}-\bm{\alpha}^*\right)^T Q\left(\bm{\alpha}-\bm{\alpha}^*\right)+\epsilon \sum_{i=1}^l\left(\alpha_i+\alpha_i^*\right)+\sum_{i=1}^l y_i\left(\alpha_i-\alpha_i^*\right) \\
        \text { subject to } & \qquad \bm{e}^T\left(\bm{\alpha}-\bm{\alpha}^*\right)=0 \\
        & 0 \leq \alpha_i, \alpha_i^* \leq C, i=1, \ldots, l,
    \end{aligned}
    \label{eq:svr_dual}
\end{equation}
where $\bm{\alpha}$ and $\bm{\alpha}^*$ are vectors of the model parameters, while $Q_{i j}$ is the kernel value for points $i$ and $j$, i.e. $K\left(\bm{x}_i, \bm{x}_j\right) \equiv \phi\left(\bm{x}_i\right)^T \phi\left(\bm{x}_j\right)$. The function solving the problem formulated in (\ref{eq:svr_dual}) is given by a~following formula

\begin{equation}
    f(x) = \sum_{i=1}^N\left(-\alpha_i+\alpha_i^*\right) K\left(\bm{x}_i, \bm{x}\right)+b.
\label{eq:svr_function}
\end{equation}
Note that the SVR model requires specifying two hyperparameters, $C$ and $\epsilon$, and a~kernel function $K(\cdot,\cdot)$. Throughout the paper, we use an arbitrary choice of $C = 1$ and $\epsilon = 0.1$, which proved to work well in the considered setting. In the simulation study, we use the scikit-learn \cite{scikit-learn} version 1.3.0 implementation of the SVR model \cite{scikit_SVR}.

\subsubsection{Specification for electricity price forecasting}
The kernel function has a~significant impact on the resulting forecasting accuracy of the SVR model. Therefore, in this section, we describe our choice of kernel functions. Arthur Jacot \cite{jacot_ntk} made a~significant step towards a~mathematical theory of neural networks. Although we do not use neural networks directly, we incorporate findings related to those models in the choice of the kernel functions. Specifically, he showed that a~neural network can be approximated by a~linearisation leading to the Neural Tangent Kernel (NTK) model. Moreover, this approximation becomes exact when the number of neurons in layers tends to infinity. Despite these theoretical results, using a~kernel regression based on fixed NTK, calculated at initialisation, does not yield accuracy close to the corresponding neural networks \cite{NTK_vs_DNN}. Various forms of finite width corrections were introduced to counter those differences in performance, see e.g. \cite{bai2020taylorized}. However, all of those methods demand iterative training and, thus, are computationally expensive.

As in our study, we intend to avoid high computational time, we search for a~solution that does not demand iterative training. Authors of \cite{label_aware_ntk} suggest that the performance gap in classification tasks is caused in part by the lack of local elasticity in kernel methods. According to \cite{Local_elasticity}, a~classifier is said to be locally elastic if its prediction at a~feature vector $x'$ is not significantly perturbed after the classifier is updated through stochastic gradient descent in another feature vector $x$, where $x$ corresponds to the same label but is dissimilar to $x'$. Based on those observations, authors of \cite{label_aware_ntk} argue that the lack of local elasticity of the limit NTK model may result from its label agnostic nature. 
To account for this, the authors of \cite{label_aware_ntk} propose a~method of adjusting the limit NTK by adding information about the labels to the kernel. The label information is incorporated by them additively in the form of labels forecasted by another model, e.g. linear regression. Their approach allows the NTK to gain local elasticity by reducing the leverage points' influence on the fitted class separation curve. The leverage points are the training points with the same label but significantly different features. Drawing from those findings, we also incorporate the separate forecast of the described variable (price) into our kernels. 

For kernel correction, we use the so-called na\"{i}ve forecast, which is a common benchmark in electricity price forecasting literature and, on the other hand, does not require any model assumptions. Usually, it is defined as the last known price. Here, the na\"{i}ve forecast of $P_{d,T}(s)$ from the moment $m$, is equal to the last published price of electricity for a given delivery:
\begin{equation}
    \widehat{P}_{d,T,m}^{\text{na\"{i}ve}} (s)=P_{d,T}(m-20).
    \label{eq:naive}
\end{equation} 
Precisely, in order to decrease the effect of training samples with close features but distant described variables 
have on the fitted function, we multiply the Laplace kernel, a~standard kernel used in kernel regression, by the Gaussian kernel based on the standardised na\"{i}ve forecast vector. We denote the SVR model estimated on such a~kernel as corrected SVR (cSVR). Note that we introduce a~modification to the standard definition of a~Laplace kernel by using a~$L2$ norm instead of a~$L1$ norm under the exponent. This decision will be further justified in this paper with a~comparison of the forecasting accuracy between models trained using both norms. For a~certain training set $\widetilde{\mathcal{S}}^i_{d,m,s}$ and vectors of the corresponding standardised features $\widetilde{\bm{x}}^{i}_{d,t,m,s}=(\widetilde{{x}}^{i1}_{d,t,m,s}, \widetilde{{x}}^{i2}_{d,t,m,s}, \dots, \widetilde{{x}}^{in_i}_{d,t,m,s})$ the kernel is defined as
\begin{equation}
    K^i(\widetilde{\bm{x}}^{i}_{d,t,m,s}, \widetilde{\bm{x}}^{i}_{d,t',m,s}) = \exp \left(-l^i_{d,m,s} \norm{\widetilde{\bm{x}}^{i}_{d,t,m,s}-\widetilde{\bm{x}}^{i}_{d,t',m,s}}_2\right)\cdot \exp \left(-g_{d,m}\norm{\widetilde{P}_{d,t}(m - 20) - \widetilde{P}_{d,t'}(m - 20)}_2^2\right)
    \label{eq:correction_of_corrected_laplace_kernel}
\end{equation}
where $t, t' \in \mathcal{T}$, $\widetilde{P}_{d,t}(m - 20)$ is the standardised na\"{i}ve forecast (see (\ref{eq:standardisation})) and $l^i_{d,m,s}$, $g_{d,m} \in \mathbb{R}_{+}$ are the kernels widths. The $L^2$ norm used in (\ref{eq:correction_of_corrected_laplace_kernel}) is defined for a~vector $\bm{x}$ with $n$ elements as $||\mathbf{x}||_2 = \sqrt{\sum_{k=1}^n\left|x_k\right|^2}$. 
The choice of a~Laplace kernel is dictated by its similarity to NTK for dense neural networks \cite{laplace_NTK_similarity}, while the Gaussian correction kernel is chosen because its central part of the distribution is flatter than the Laplace one. Such choice allows avoiding overly correcting the Laplace kernel for close feature values. After the correction of the Laplace kernel with the Gaussian kernel, the only high values of $K(\widetilde{\bm{x}}^{i}_{d,t,m,s}, \widetilde{\bm{x}}^{i}_{d,t',m,s})$ are those where both the na\"{i}ve forecasts and variables sets from days $t$ and $t'$ are close. In other words, if the samples on days $t$ and $t'$ are distant in terms of $\widetilde{\bm{x}}^{i}_{d,t,m,s}$ and $\widetilde{\bm{x}}^{i}_{d,t',m,s}$, the na\"{i}ve forecast in both days needs to be close for the kernel value to avoid further reduction. Otherwise, the value of the Laplace and Gaussian kernels' product will be small, thus decreasing the importance of pair $(t, t')$ in the resulting fit; see equation (\ref{eq:svr_function}). Note that this effect is only partially in line with the notion of local elasticity, which, formulated for the classification, demands that we limit the influence of point $t$ on the function at point $t'$ if $\widetilde{\bm{x}}^{i}_{d,t,m,s}$ is distant from $\widetilde{\bm{x}}^{i}_{d,t',m,s}$ and the label of both features vectors is the same. In our kernel modification, we limit the influence of point $t$ on the function at point $t'$ if na\"{i}ve forecasts from those days are distant. Thus, the modification of the kernel reduces the number of points influencing the fit. 

Using correction (\ref{eq:correction_of_corrected_laplace_kernel}) and the considered training data (see Section \ref{sec:expert_sets} for details), we suggest three kernels to be used in the forecasting task. Each corresponds to variables set $\widetilde{\mathcal{S}}^i_{d,m,s}$, $i=1,2,3$ and the formula (\ref{eq:correction_of_corrected_laplace_kernel}). Let us denote them as $K^i$. 
To evaluate the result of correcting the Laplace kernel with the Gaussian kernel based on the na\"{i}ve forecast, we also test the performance of the standalone Laplace kernel $K^4$, built using the $\widetilde{\mathcal{S}}^1_{d,m,s}$ variables set. Recall that it is the set consisting of all explanatory variables considered in the study. Such a~Laplace kernel is given by

\begin{equation}
    K^{4}(\widetilde{\bm{x}}^{1}_{d,t,m,s}, \widetilde{\bm{x}}^{1}_{d,t^{'},m,s}) = \exp \left(-l^1_{d,m,s}\norm{\widetilde{\bm{x}}^{1}_{d,t,m,s}-\widetilde{\bm{x}}^{1}_{d,t^{'},m,s}}_2\right).
    \label{eq:kernel_K4}
\end{equation}
To justify our choice of $L2$ norm instead of $L1$ norm in the Laplace kernel, we compare the forecasting accuracy of the SVR model estimated using both versions for the set $\widetilde{\mathcal{S}}^1_{d,m,s}$. Thus, we define also the Laplace kernel with the $L1$ norm as $K^5$
\begin{equation}
    K^{5}(\widetilde{\bm{x}}^{1}_{d,t,m,s}, \widetilde{\bm{x}}^{1}_{d,t^{'},m,s}) = \exp \left(-l^{1'}_{d,m,s}\norm{\widetilde{\bm{x}}^{1}_{d,t,m,s}-\widetilde{\bm{x}}^{1}_{d,t^{'},m,s}}\right).
    \label{eq:kernel_K5}
\end{equation}

The kernel widths $l^i_{d,m,s}$ and $g_{d,m}$ are calculated based on the distribution of the input data. This solution is inspired by \cite{kernel_width_based_on_median}, where the authors set the scaling parameter of the Gaussian kernel to be equal to the median Euclidean distance between the training samples. We incorporate a~similar algorithm in our study as it allows us to choose the widths based on the characteristics of the sample without performing computationally expensive cross-validation. A~key observation that allows us to choose kernel widths based on data quantile is that the Laplace and Gaussian kernels correspond to
probability density functions of those distributions. Therefore, setting the weight of a~kernel such that a~certain amount of distances between samples falls below a~set quantile of a distribution is a~straightforward task. In the case of the Laplace kernel, we arbitrarily set the third quartile of the Laplace distribution to the value of the median of the kernel arguments. Recall that 
the quantile function of the Laplace distribution for $p \geq 0.5$, i.e. above the median, is

\begin{equation}
    Q(p) = \mu - b\cdot \ln{\left( 2 - 2p \right)},
    \label{eq:laplace_quantile_func}
\end{equation}
where $\mu$ and $b$ are the distribution parameters. Let us now set the location parameter $\mu$ to $0$ and recall that the width of the resulting Laplace kernel is equal to $\frac{1}{b}$. Let us define the $\hat{q}^{i}_{d,m,s}(0.5)$ as the empirical median of $L2$ distances set $\Bigl\{ \norm{\widetilde{\bm{x}}^{1}_{d,t,m,s}-\widetilde{\bm{x}}^{1}_{d,t^{'},m,s}}_2$, $t,t'\in \mathcal{T} \Bigr\}$. Then, after putting $p = 0.75$ 
 and comparing (\ref{eq:laplace_quantile_func}) to the empirical median $\hat{q}^{i}_{d,m,s}(0.5)$ we get
\begin{equation}
    \hat{q}^{i}_{d,m,s}(0.5) = -\frac{1}{l^i_{d,m,s}}\ln{(2 - 2\cdot0.75)},
    \label{eq:laplace_width_problem}
\end{equation}
which gives the equation for the Laplace kernel width, which guarantees that half of the closest training samples fall below the third quartile of the Laplace distribution corresponding to this kernel:
\begin{equation}
    l^i_{d,m,s} = \left|\frac{-\ln{(2 - 2\cdot0.75)}}{\hat{q}^{i}_{d,m,s}(0.5)}\right|.
    \label{eq:width_laplace}
\end{equation}
Note here that the absolute value in (\ref{eq:width_laplace}) is put in order to ensure the positive value of the kernel widths. Note also that the $l^{1'}_{d,m,s}$ in (\ref{eq:kernel_K5}) is calculated in a~similar way, the only difference being the empirical median, which is calculated from the $L1$ distances set. 

To derive $g_{d,m}$ in a similar way, we set the third quartile of the Gaussian distribution equal to the third quartile of the kernel correction input. Recall that 
quantile function of the Gaussian distribution is
\begin{equation}
    Q(p) = \mu + \sigma\sqrt{2}erf^{-1}(2p - 1).
\end{equation}
Setting the location parameter $\mu$ to $0$ and substituting the quantile of the standard normal distribution, $z_p$, for $\sqrt{2}erf^{-1}(2p - 1)$, we derive the desired width of the Gaussian kernel
\begin{equation}
    g_{d,m} = \frac{1}{2\sigma_{d,m}^2}
    \label{eq:width_gaussian}
\end{equation}
with
\begin{equation}
    \sigma_{d,m} = \frac{\hat{q}_{d,m}(0.75)}{z_{0.75}}\label{eq:sigma},
\end{equation}
where $\hat{q}_{d,m}(0.75)$ is the empirical third quartile of the squared $L2$ distances set $\Bigl\{ \norm{\widetilde{P}_{d,t}(m - 20) - \widetilde{P}_{d,t^{'}}(m - 20)}_2^2$, $t, t' \in \mathcal{T} \Bigr \}$. Note that, although such a~definition can be extended freely to other combinations of empirical and theoretical percentiles, $z_{0.5}$, i.e. the median, cannot be used as it leads to the division by zero in (\ref{eq:sigma}).

\begin{figure}[!h]
    \centering
    \includegraphics[width=0.7\textwidth]{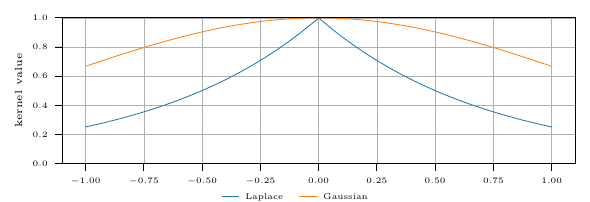}
    \caption{Kernels with quantile dependent widths, (\ref{eq:width_laplace}), (\ref{eq:width_gaussian}).}
    \label{fig:Laplace_and_Gaussian_kernels_example}
\end{figure}

The definitions \ref{eq:width_laplace} and \ref{eq:width_gaussian} of the kernel widths allow one to control the kernel values, which now are contained within predictable boundaries. The relation between the Gaussian and Laplace kernels, derived using the widths (\ref{eq:width_gaussian}) and (\ref{eq:width_laplace}), on synthetic arguments spaced equally from $-1$ to $1$, is shown in Figure \ref{fig:Laplace_and_Gaussian_kernels_example}. The Laplace kernel is much tighter for the same inputs, as we put the empirical median in the theoretical third quartile. This ensures that half of the input values lie outside the third quartile of the Gaussian density related to such kernel.

To finish the description of the SVR model for the intraday electricity market, let us define the corresponding price forecast in terms of kernels $K^k\left(\widetilde{\bm{x}}^{i}_{d,t,m,s}, \widetilde{\bm{x}}^{i}_{d,T,m,s}\right)$, where $k$ denotes the kernel type, while $i$ the explanatory variables set, i.e. here we have $(k, i)=(1,1)$, $(2,2)$, $(3,3)$, $(4,1)$ $(5, 1)$. Using Equation (\ref{eq:svr_function}) the differenced standardised price forecast can be written as

\begin{equation}
    \widehat{\widetilde{P\,}}\!{'}^i_{d,T,m}\left(s\right) = \sum_{t\in\mathcal{T}}\left(-\alpha_t+\alpha_t^*\right) K^k\left(\widetilde{\bm{x}}^{i}_{d,t,m,s}, \widetilde{\bm{x}}^{i}_{d,T,m,s}\right)+b,
\label{eq:svr_prediction}
\end{equation}
where $\alpha_t$, $\alpha_t^*$ and $b$ are the model parameters and $\widetilde{\bm{x}}^{i}_{d,t,m,s}$ are the vectors of explanatory variables from a training window of length $n$.

\subsection{LASSO}

We compare the cSVR model performance with the LASSO, which is a~state-of-the-art model in the field of electricity price forecasting. Its good performance was shown e.g. by \cite{UNIEJEWSKI2019}, where the authors used LASSO to forecast the closing value of ID3 of the German continuous intraday market. The time of the forecast used by the authors was set to 4 hours before the delivery. Although this approach cannot be translated directly to our case, we test the LASSO performance, treating it as a~benchmark model.

In the LASSO approach, the price is assumed to be given by the~following linear combination
\begin{equation}
    \widetilde{P\,}\!{'}^i_{d,t,m}(s) = \sum_{j = 1}^{n_i} \beta^{i,j}_{d,m,s}\widetilde{\bm{x}}^{i,j}_{d,t,m,s} + \epsilon^i_{d,t,m,s}
    \label{eq:lasso_model}
\end{equation}
where $\widetilde{\bm{x}}^{i,j}_{d,t,m,s}$ are the explanatory variables from set $\widetilde{\mathcal{S}}^i_{d,m,s}$, and $n_i$ is the total number of variables in that set. The estimation of the parameters $\beta^{i,j}_{d,m,s}$ is based on the penalized minimization of the residual sum of squares (RSS)~\cite{lasso}, i.e. finding $\min_{\bm{\beta}^i_{d,m,s}}\left\{\mathrm{RSS}+\lambda \sum_{j=1}^{n_i}\left|\beta^{i,j}_{d,m,s}\right|\right\}$. The forecast of the differenced, standardised price in the LASSO model can be written using equation (\ref{eq:lasso_model}) as $\widehat{\widetilde{P\,}}\!{'}^i_{d,T,m}(s) = \sum_{j = 1}^{n_i} \hat{\beta}^{i,j}_{d,m,s}\widetilde{\bm{x}}^{i,j}_{d,T,m,s}$.
The model regularisation parameter $\lambda$ is updated at each forecasting point separately. We choose the cross-validation based on Least Angle Regression (LARS) \cite{efron_lars} as the algorithm for finding the optimal $\lambda$. We run the LASSO LARS algorithm 14 times in an expanding window, leave-one-out, cross-validation scheme. The first window is cut from the whole training window used to train the LASSO model by leaving the last 14 days out. All consecutive windows are obtained by adding one day iteratively. LARS allows finding the entire path of the LASSO solution along $\lambda$ without the need for specifying the range of $\lambda$s to be tested. As a~result of cross-validation, the regularisation parameter $\lambda$ giving the lowest mean squared error (MSE) is chosen. Using this parameter, we train the LASSO LARS model on the whole training window and forecast the out-of-sample prices. The implementation used in this study is scikit-learn LassoLarsCV \cite{scikit_LASSOLARS}.
\label{sec:lasso_limit_variables}

Since in the parametric LASSO model the number of training variables should not exceed the number of training points, we reapply the correlation-based filtering before training the LASSO. Since in our application only the size of $\widetilde{\mathcal{S}}^1_{d,m,s}$ exceeds the length of the training sample, we apply the filter only to this set. We recursively exclude the variables from pairs with the highest correlation until the number of variables is less than the training window length reduced by 14, i.e. the length of the $\lambda$ calibration window. Again, the correlation filter is not applied to the exogenous variables.

\subsection{Random Forest}

A second reference model that we choose for the comparison with cSVR is the Random Forest (RF). It is an ensemble of decision tree algorithms, introduced by \cite{random_forest}. It involves constructing a~large number of decision trees from bootstrap samples from the training dataset, like bagging. However, unlike bagging, RF also involves selecting a~subset of input variables at each split point in the construction of the trees. RF prediction is then chosen as the average prediction across the decision trees. Due to the size of the forecasting study and its computational time, we set the number of trees at $256$ and their maximum depth at $8$. We use the scikit-learn implementation of RF \cite{scikit_RF}. We acknowledge the limitation of setting both parameters arbitrarily. However, the computational time required to systematically find the optimal values would exceed our computational capacity in a~given task. The chosen values are, hence, a trade-off between computational time and expected forecasting accuracy.

We use three previously defined input sets to RF, namely $\widetilde{\mathcal{S}}^i_{d,m,s}$, $i=1$,$2$,$3$. In contrast to LASSO, RF is a~non-parametric model, which allows the number of variables to vastly exceed the number of training points. Therefore, we do not use the second iteration of the correlation filtering, as in the case of LASSO for $\widetilde{\mathcal{S}}^1_{d,m,s}$ variables set.

\section{Electricity price forecasting}
\label{sec:forecasting_study}
In the case of continuous intraday market transactions on electricity delivered within a given quarter hour $d$ on day $T$ can be settled from 16:00 on the day prior to the delivery up to 5 minutes before. Therefore, we need to consider an additional time dimension, namely trade time $s$. The developed algorithm needs to be able at time $m$ to forecast the price $P_{d,T}(s)$ of transaction at minute $s$, for delivery in quarter hour $d$ and day $T$, where $m<s<d$ (see Figure \ref{fig:forecasting_timeline} for a schematic illustration of the considered timelines). This differs from the standard auction-based markets setup, where there is only one time point of transaction settlement. 
Testing the model performance on such a~high number of possibilities is not computationally feasible. Therefore, we set  a~range of representative horizons for forecasting and trade times. To investigate the models performance over the whole day, we consider all of quarter-hourly deliveries. The lead times, i.e. times from transaction to delivery, are based on observations from data analysis performed in Section \ref{sec:dataset}. The most active period on the market is between 3 hours and 30 minutes before delivery. Hence, we take into account six, equally distant, lead times from 180 to 30 minutes. As for forecasting horizon, we take every 30 minutes from 30 to 210 minutes horizons and then every 90 minutes from 210 to 480 minutes horizons. This gives in total 5760 forecast calculations for all these cases within a single day. 

We calculate the price forecasts for each of the sets of explanatory variables $\widetilde{\mathcal{S}}^i_{d,m,s}, i=1,2,3$ as well as the {na\"{i}ve} forecasts treated further as a~benchmark. In order to obtain the final price forecast from the cSVR, SVR, LASSO and RF models, both standardisation and differencing steps need to be inverted, i.e. 
\begin{equation}
    \widehat{P}^i_{d,T,m}(s) = \sigma_{d,s}\widehat{\widetilde{P\,}}\!{'}^i_{d,T,m}(s) + \mu_{d,s} + P_{d,T}(m - 20).
\end{equation} 

Moreover, we apply a forecast averaging over the three considered expert sets. In recent years, forecast averaging has become a state-of-the-art approach to increase the accuracy of electricity price predictions, given multiple forecasts with different characteristics; see, e.g. \cite{forecasts_combination_weron_nowotarski, maciejo_PCA_averaging}. We compare two averaging schemes. First is the simple arithmetic average, which was used with success by \cite{Hubicka2019ANO} and \cite{forecasts_combination_weron_nowotarski} for the day-ahead market~prices:
\begin{equation}
    \widehat{P}^{\text{avg.}}_{d,T,m}(s)  = \frac{\widehat{P}^1_{d,T,m}(s)  +\widehat{P}^2_{d,T,m}(s)  + \widehat{P}^3_{d,T,m}(s)  + \widehat{P}^{\text{na\"{i}ve}}_{d,T,m}(s) }{4}.
    \label{eq:average}
\end{equation}
The second averaging scheme used in this paper takes into account the out-of-sample errors of each of the considered model forecasts. Following \cite{averaging_forecasts} the averaged forecast is given by
\begin{equation}
    \begin{split}
        &\widehat{P}^{\text{w. avg.}}_{d,T,m}(s)  = w^1_W\widehat{P}^1_{d,T,m}(s)  + w^2_W\widehat{P}^2_{d,T,m}(s)  + w^3_W\widehat{P}^3_{d,T,m}(s)  + w^4_W\widehat{P}^{\text{na\"{i}ve}}_{d,T,m}(s) , \\&
        w^j_W = \frac{\frac{1}{\mathrm{MAE}^j_W}}{\sum_{j = 1}^4 \frac{1}{\mathrm{MAE}^j_W}},
    \end{split}
    \label{eq:weighted_avg}
\end{equation}
where $\widehat{P}^{\text{w. avg.}}_{d,T,m}(s) $ is the resulting weighted average price forecast and $W$ is the length of the window on which the out-of-sample mean absolute error (MAE) is calculated. Here, it is equal to 7 days. The authors of \cite{averaging_forecasts} showed that such a~weighted average outperforms the averaging with equal weights (\ref{eq:average}).

\begin{figure}[!ht]
\centering
\resizebox{1\textwidth}{!}{%
\begin{circuitikz}
\tikzstyle{every node}=[font=\normalsize]
\draw [rounded corners = 6.0] (0,23.75) rectangle  node {\normalsize Market data} (15.5,22.75);
\draw [ fill={rgb,255:red,213; green,232; blue,212} , rounded corners = 6.0, ] (0,16.5) rectangle  node {\normalsize na\"{\i}ve} (3.5,17.25);
\draw [ fill={rgb,255:red,218; green,232; blue,252} , rounded corners = 6.0, ] (4,22.25) rectangle  node {\normalsize $\widetilde{S}^1$} (5,21.5);
\draw [ fill={rgb,255:red,218; green,232; blue,252} , rounded corners = 6.0, ] (5.25,22.25) rectangle  node {\normalsize $\widetilde{S}^2$} (6.25,21.5);
\draw [ fill={rgb,255:red,218; green,232; blue,252} , rounded corners = 6.0, ] (6.5,22.25) rectangle  node {\normalsize $\widetilde{S}^3$} (7.5,21.5);
\draw [ fill={rgb,255:red,218; green,232; blue,252} , rounded corners = 6.0, ] (8,22.25) rectangle  node {\normalsize $\widetilde{S}^1$} (9,21.5);
\draw [ fill={rgb,255:red,218; green,232; blue,252} , rounded corners = 6.0, ] (9.25,22.25) rectangle  node {\normalsize $\widetilde{S}^2$} (10.25,21.5);
\draw [rounded corners = 0.0] (9.5,22) rectangle (9.5,22);
\draw [ fill={rgb,255:red,218; green,232; blue,252} , rounded corners = 6.0, ] (10.5,22.25) rectangle  node {\normalsize $\widetilde{S}^3$} (11.5,21.5);
\draw [ fill={rgb,255:red,218; green,232; blue,252} , rounded corners = 6.0, ] (12,22.25) rectangle  node {\normalsize $\widetilde{S}^1$} (13,21.5);
\draw [ fill={rgb,255:red,218; green,232; blue,252} , rounded corners = 6.0, ] (13.25,22.25) rectangle  node {\normalsize $\widetilde{S}^2$} (14.25,21.5);
\draw [ fill={rgb,255:red,218; green,232; blue,252} , rounded corners = 6.0, ] (14.5,22.25) rectangle  node {\normalsize $\widetilde{S}^3$} (15.5,21.5);
\draw [ fill={rgb,255:red,248; green,206; blue,204} , rounded corners = 6.0, ] (4,19) rectangle  node {\normalsize cSVR} (7.5,19.75);
\draw [ fill={rgb,255:red,225; green,213; blue,231} , rounded corners = 6.0, ] (8,19) rectangle  node {\normalsize LASSO} (11.5,19.75);
\draw [ fill={rgb,255:red,255; green,230; blue,204} , rounded corners = 6.0, ] (12,19) rectangle  node {\normalsize RF} (15.5,19.75);
\draw [ fill={rgb,255:red,248; green,206; blue,204} , rounded corners = 6.0, ] (4,18.5) rectangle  node {\normalsize $\widehat{P}^1$} (5,17.75);
\draw [ fill={rgb,255:red,248; green,206; blue,204} , rounded corners = 6.0, ] (5.25,18.5) rectangle  node {\normalsize $\widehat{P}^2$} (6.25,17.75);
\draw [ fill={rgb,255:red,248; green,206; blue,204} , rounded corners = 6.0, ] (6.5,18.5) rectangle  node {\normalsize $\widehat{P}^3$} (7.5,17.75);
\draw [ fill={rgb,255:red,225; green,213; blue,231} , rounded corners = 6.0, ] (8,18.5) rectangle  node {\normalsize $\widehat{P}^1$} (9,17.75);
\draw [ fill={rgb,255:red,225; green,213; blue,231} , rounded corners = 6.0, ] (9.25,17.75) rectangle  node {\normalsize $\widehat{P}^2$} (10.25,18.5);
\draw [ fill={rgb,255:red,225; green,213; blue,231} , rounded corners = 6.0, ] (10.5,17.75) rectangle  node {\normalsize $\widehat{P}^3$} (11.5,18.5);
\draw [ fill={rgb,255:red,255; green,230; blue,204} , rounded corners = 6.0, ] (12,18.5) rectangle  node {\normalsize $\widehat{P}^1$} (13,17.75);
\draw [ fill={rgb,255:red,255; green,230; blue,204} , rounded corners = 6.0, ] (13.25,17.75) rectangle  node {\normalsize $\widehat{P}^2$} (14.25,18.5);
\draw [ fill={rgb,255:red,255; green,230; blue,204} , rounded corners = 6.0, ] (14.5,17.75) rectangle  node {\normalsize $\widehat{P}^3$} (15.5,18.5);
\draw [ fill={rgb,255:red,248; green,206; blue,204} , rounded corners = 6.0, ] (5.25,17.25) rectangle  node {\normalsize $P^{\text{avg.}}$} (6.25,16.5);
\draw [ fill={rgb,255:red,225; green,213; blue,231} , rounded corners = 6.0, ] (9.25,16.5) rectangle  node {\normalsize $P^{\text{avg.}}$} (10.25,17.25);
\draw [ fill={rgb,255:red,255; green,230; blue,204} , rounded corners = 6.0, ] (13.25,16.5) rectangle  node {\normalsize $P^{\text{avg.}}$} (14.25,17.25);
\draw [->, >=Stealth] (1.75,22.75) -- (1.75,17.25);
\draw [ fill={rgb,255:red,225; green,213; blue,231} , rounded corners = 6.0, ] (8,21.25) rectangle  node {\normalsize filter} (9,20); 
\draw [line width=0.3pt, ->, >=Stealth] (4.5,22.75) -- (4.5,22.25);
\draw [line width=0.3pt, ->, >=Stealth] (5.75,22.75) -- (5.75,22.25);
\draw [line width=0.3pt, ->, >=Stealth] (7,22.75) -- (7,22.25);
\draw [line width=0.3pt, ->, >=Stealth] (8.5,22.75) -- (8.5,22.25);
\draw [line width=0.3pt, ->, >=Stealth] (9.75,22.75) -- (9.75,22.25);
\draw [line width=0.3pt, ->, >=Stealth] (11,22.75) -- (11,22.25);
\draw [line width=0.3pt, ->, >=Stealth] (12.5,22.75) -- (12.5,22.25);
\draw [line width=0.3pt, ->, >=Stealth] (13.75,22.75) -- (13.75,22.25);
\draw [line width=0.3pt, ->, >=Stealth] (15,22.75) -- (15,22.25);
\draw [line width=0.3pt, ->, >=Stealth] (8.5,21.5) -- (8.5,21.25);
\draw [line width=0.3pt, ->, >=Stealth] (8.5,20) -- (8.5,19.75);
\draw [line width=0.3pt, ->, >=Stealth] (4.5,21.5) -- (4.5,19.75);
\draw [line width=0.3pt, ->, >=Stealth] (5.75,21.5) -- (5.75,19.75);
\draw [line width=0.3pt, ->, >=Stealth] (7,21.5) -- (7,19.75);
\draw [line width=0.3pt, ->, >=Stealth] (4.5,19) -- (4.5,18.5);
\draw [line width=0.3pt, ->, >=Stealth] (5.75,19) -- (5.75,18.5);
\draw [line width=0.3pt, ->, >=Stealth] (7,19) -- (7,18.5);
\draw [line width=0.3pt, ->, >=Stealth] (4.5,17.75) -- (5.75,17.25);
\draw [line width=0.3pt, ->, >=Stealth] (5.75,17.75) -- (5.75,17.25);
\draw [line width=0.3pt, ->, >=Stealth] (7,17.75) -- (5.75,17.25);
\draw [line width=0.3pt, ->, >=Stealth] (8.5,17.75) -- (9.75,17.25);
\draw [line width=0.3pt, ->, >=Stealth] (9.75,17.75) -- (9.75,17.25);
\draw [line width=0.3pt, ->, >=Stealth] (11,17.75) -- (9.75,17.25);
\draw [line width=0.3pt, ->, >=Stealth] (12.5,17.75) -- (13.75,17.25);
\draw [line width=0.3pt, ->, >=Stealth] (13.75,17.75) -- (13.75,17.25);
\draw [line width=0.3pt, ->, >=Stealth] (15,17.75) -- (13.75,17.25);
\draw [line width=0.3pt, ->, >=Stealth] (12.5,19) -- (12.5,18.5);
\draw [line width=0.3pt, ->, >=Stealth] (8.5,19) -- (8.5,18.5);
\draw [line width=0.3pt, ->, >=Stealth] (9.75,19) -- (9.75,18.5);
\draw [line width=0.3pt, ->, >=Stealth] (11,19) -- (11,18.5);
\draw [line width=0.3pt, ->, >=Stealth] (13.75,19) -- (13.75,18.5);
\draw [line width=0.3pt, ->, >=Stealth] (15,19) -- (15,18.5);
\draw [line width=0.3pt, ->, >=Stealth] (9.75,21.5) -- (9.75,19.75);
\draw [line width=0.3pt, ->, >=Stealth] (11,21.5) -- (11,19.75);
\draw [line width=0.3pt, ->, >=Stealth] (12.5,21.5) -- (12.5,19.75);
\draw [line width=0.3pt, ->, >=Stealth] (13.75,21.5) -- (13.75,19.75);
\draw [line width=0.3pt, ->, >=Stealth] (15,21.5) -- (15,19.75);
\end{circuitikz}
}%
\caption{Flowchart of the forecast calculation for each of the considered models. Note that, for clarity, the indexes are omitted in the notation.}
\label{fig:flowchart_of_models}
\end{figure}
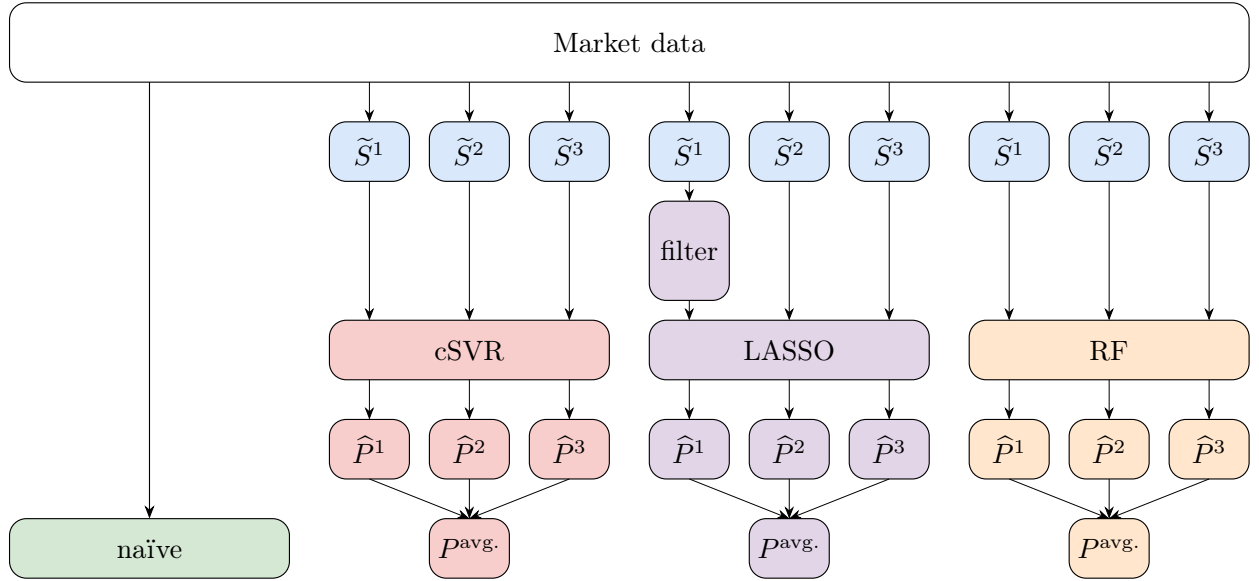

Each forecast computation is performed in an expanding window scheme. The first training window ranges from 1 November 2018 to 24 December 2019. Models trained on this window are used for forecasting the prices at 25 December 2019. A~schematic illustration of the entire case study flowchart is shown in Figure \ref{fig:flowchart_of_models}. We start with building expert sets of standardised explanatory variables $\widetilde{\mathcal{S}}^i_{d,m,s}$, $i=1,2,3$ (as in (\ref{eq:S1})-(\ref{eq:S3})). We then train each model separately on those sets. Recall here that an additional step is performed for the LASSO model and set $\widetilde{\mathcal{S}}^1_{d,m,s}$, where we apply a second correlation filtering to limit the number of variables (see Section \ref{sec:lasso_limit_variables}). We repeat this process 372 times, each time expanding the training window by one day to forecast prices for the remaining days in 2019 and for all days in  2020. Then, using the forecasts from three sets $\widetilde{\mathcal{S}}^i_{d,m,s}, i=1,2,3$ and the na\"{i}ve, we calculate their average according to formula (\ref{eq:average}) or (\ref{eq:weighted_avg}). The first seven days of forecasts, i.e. 24 to 31 December 2019, are used as a~calibration window to tune the averaging weights for 1 January 2020. Then, when the window is shifted, the days from 25 December 2019 to 1 January 2020 are used to calibrate the weights for 2 January 2020 and so on.

\subsection{Forecasting results}
The forecasting accuracy of the models considered is evaluated in a yearly test window, which stretches from 01.01.2020 to 31.12.2020. For the error measure, we use MAE:
\begin{equation}
    \mathrm{MAE}^{\text{model}^\text{variant}}_{d,m,s} = \frac{1}{N}\sum_{j=1}^N \left| \widehat{P}^\text{variant}_{d,j,m}(s) - P_{d,j}(s) \right|,
    \label{eq:MAE}
\end{equation}
where $\widehat{P}^\text{variant}_{d,j,m}(s)$ is the forecast from a~given model and variant. A~model might be the na\"{i}ve, cSVR, SVR, LASSO or RF, see the flowchart in Figure \ref{fig:flowchart_of_models}. The variant for the SVR and cSVR models is tied to the kernel index, i.e. $\text{variant}=1$,$2$,$3$ for cSVR and $\text{variant}=4$,$5$ for SVR, or to the averaging schemes. For both LASSO and RF models, we will use only the w. avg. variant later in the text. $P_{d,j}(s)$ is the real price for day $j$,~delivery $d$~and transaction time $s$,~while $N$~is the length of the test window. 

As a~benchmark for all of the considered models, we use the na\"{i}ve forecast, defined in (\ref{eq:naive}), i.e. the last known price. It is a common approach in a~very short-term electricity price forecasting literature, see e.g. \cite{NARAJEWSKI2020100107, beating_the_naive}. To evaluate the improvement in forecast accuracy with respect to the na\"{i}ve forecast, we use the relative mean absolute error (rMAE). It is defined as
\begin{equation}
\mathrm{rMAE}^{\text{model}^\text{variant}}_{d,m,s} = \frac{\mathrm{MAE}^{\text{model}^\text{variant}}_{d,m,s} - \mathrm{MAE}^{\text{na\"{i}ve}}_{d,m,s}}{\mathrm{MAE}^{\text{na\"{i}ve}}_{d,m,s}}.
\label{eq:relative_MAE}
\end{equation}
 Note that a negative value of $\mathrm{rMAE}^{\text{model}^\text{variant}}_{d,m,s}$ means that the  error of the considered model is lower than the error of the na\"{i}ve forecast, so it indicates improvement in the forecast accuracy.

We start with an analysis of the relative accuracy of the forecasts calculated using all the considered models, i.e. cSVR, SVR, LASSO, RF, for the ~60 minutes lead time. It is chosen for this in-depth analysis due to its placement at the Single IntraDay Coupling (SIDC) closing time. The market characteristics visible in Figure \ref{fig:std_3D} suggest that a~lead time close to 60 minutes is a~time point before the rapid increase of volatility, but already with higher liquidity. The authors of \cite{Hirsh_simulation_based} also showed that volatility increases with the closure of SIDC and the time to delivery. 

\begin{figure}[!h]
    \centering
    \includegraphics[width=\textwidth]{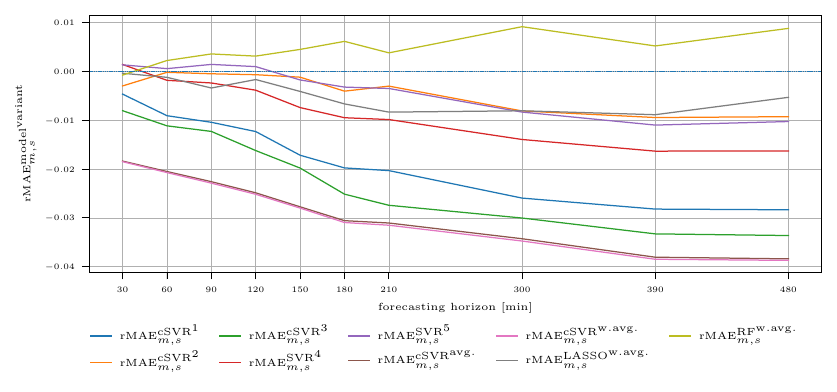}
    \caption{Relative errors, $\mathrm{rMAE}^{\text{model}^{\text{variant}}}_{m,s}$ obtained for forecasts calculated using different approaches: cSVR calculated for individual variables sets cSVR$^i, i=1,2,3$; averaged cSVR, cSVR$^{\text{avg.}}$ and cSVR$^{\text{w.avg.}}$; SVR with $L1$ and $L2$ norm, SVR$^{4}$ and SVR$^{5}$; weighted averages of the LASSO and RF forecasts, LASSO$^{\text{w.avg}}$ and RF$^{\text{w.avg.}}$.  The errors were averaged over all 96 deliveries $d$. Lead time is equal to 60 minutes.} 
    \label{fig:relative_accuracy}
\end{figure}

The obtained results, averaged across all 96 deliveries $\mathrm{rMAE}^{\text{model}^\text{variant}}_{m,s}=\frac{1}{96}\sum_{d=1}^{96} \mathrm{rMAE}^{\text{model}^\text{variant}}_{d,m,s}$, are presented in Figure \ref{fig:relative_accuracy}. 
 The figure contains the relative errors for the models considered and their modifications, namely the cSVR results for all the variables sets, $\text{rMAE}_{m,s}^{\text{cSVR}^i}$ $i=1,2,3$; their forecast averages $\text{rMAE}_{m,s}^{\text{cSVR}^{\text{avg.}}}$ and $\text{rMAE}_{m,s}^{\text{cSVR}^{\text{w. avg.}}}$; the SVR model results for the Laplace kernel with $L1$ norm $\text{rMAE}_{m,s}^{\text{SVR}^5}$ and with $L2$ norm $\text{rMAE}_{m,s}^{\text{SVR}^4}$; and finally, the best performing LASSO and RF models, $\text{rMAE}_{m,s}^{\text{LASSO}^{\text{w. avg.}}}$ and $\text{rMAE}_{m,s}^{\text{RF}^{\text{w. avg.}}}$.  Let us first analyse the results of the proposed cSVR model trained on three separate variables sets $\widetilde{\mathcal{S}}^i_{d,m,s}, i=1,2,3$. The best results are obtained using the set which contains only the exogenous variables and the last information about the price, i.e. $\widetilde{\mathcal{S}}^3_{d,m,s}$. The model trained on $\widetilde{\mathcal{S}}^1_{d,m,s}$, i.e. the training set containing broad information about the price trajectory and exogenous variables, yields slightly worse results. However, in both cases, the relative errors are similar in magnitude and shape. Moreover, they are decreasing with the forecast horizon, which is caused by the fact that the  na\"{i}ve forecast is losing its accuracy with increasing distance from the time of trade. The worst performing cSVR variant is trained on $\widetilde{\mathcal{S}}^2_{d,m,s}$, i.e. the close information set containing only the minutely sampled prices from the last hour. However, both the weighted and arithmetic averages of the cSVR model forecasts corresponding to all three sets and the na\"{i}ve, outperform the na\"{i}ve and the other models. The difference between arithmetic and the weighted averaging scheme is not significant here. 

Next, we analyse the effect of the proposed kernel correction on the forecast accuracy. The relative forecast errors obtained for the set $\widetilde{\mathcal{S}}^1_{d,m,s}$ and the SVR model without kernel correction, $K^4$ and $K^5$, (\ref{eq:kernel_K4})-(\ref{eq:kernel_K5}), are also plotted in Figure \ref{fig:relative_accuracy}. For comparison, we use two norms in the Laplace kernel: the $L1$ norm, as it is done classically in the literature, as well as the $L2$ norm, which is consistent with the Gaussian kernel correction. First, note that applying the $L2$ norm in the Laplace kernel proves to perform better than applying the same kernel with the $L1$ norm. Hence, further, we use the $L2$ norm for all Laplace kernels. Second, the performance of the standard SVR, i.e. without kernel correction, is significantly worse than the proposed cSVR model with kernel corrected by the na\"{i}ve forecast. 

 Finally, both the LASSO and RF models perform worse than the cSVR. The difference between relative LASSO and cSVR errors is close to $2\%$ up to 200 minutes horizon and increases up to $3.5\%$ for 480 minutes horizon. Similarly, the considered RF model performs worse than the na\"{i}ve for all of the forecasting horizons except for the 30 minutes one. Thus, the difference in relative error between the RF and cSVR models is up to almost $5\%$ for the 480 minutes horizon. Although the LASSO is better than the na\"{i}ve for forecasting horizons over 120 minutes, it is still significantly less accurate than the averaged cSVR models. Interestingly, for both the LASSO and RF models, the relative errors are larger for longer forecasting horizons, while for the cSVR model, they tend to decrease with the forecasting horizon. 
For clarity, in Figure \ref{fig:relative_accuracy} we plot the LASSO and RF errors obtained only with the forecast averaging scheme, as these were the most accurate results. The corresponding relative errors for different sets of variables are also presented in Table \ref{tab:avg_results}. Here, the values are averaged over all forecast horizons. The relative error of the weighted averaged cSVR forecast is approximately $3\%$ lower than the na\"{i}ve. For the other models, the only improvement is obtained for the average LASSO and is approximately $0.5\%$. For all other cases, the mean errors are higher than in the case of the na\"{i}ve forecast. In general, the analysis conducted for 60 minutes lead time shows that the proposed cSVR approach outperforms LASSO and RF in the considered cases. 
 What is interesting is that the LASSO estimated on the set $\widetilde{\mathcal{S}}^1_{d,m,s}$ performs poorly compared to $\widetilde{\mathcal{S}}^2_{d,m,s}$, which was not the case for cSVR. This shows that even after reapplying the correlation filter, the LASSO model is not coping well with the number of explanatory variables close to the length of the training sample.

\begin{figure}[!h]
    \centering
    \includegraphics[width=\textwidth]{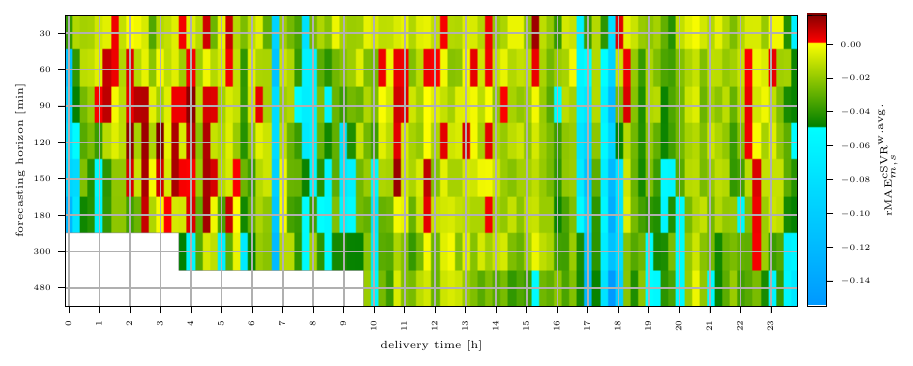}
    \caption{Relative MAE, (\ref{eq:relative_MAE}), for the cSVR model forecasts averaged as in (\ref{eq:weighted_avg}) and lead time of 60 minutes. Deliveries for which the cSVR model on a~given forecasting horizon performed worse than the na\"{i}ve are marked with red color, while cases for which the cSVR model performed better by over $5\%$ than the na\"{i}ve are marked with blue color.}
    \label{fig:3D_relative_MAE}
\end{figure}

\begin{table}[!h]
\centering
\caption{Relative errors $r\mathrm{MAE}^{model^{variant}}_{s}$ averaged over deliveries  and forecast horizons for lead time equal to 60 minutes. The value for the best performing approach is marked with bold.}
\begin{tabular}{|l|c|cccc|cccc|}
\hline
model &
  cSVR &
  \multicolumn{4}{c|}{LASSO} &
  \multicolumn{4}{c|}{RF} \\ \hline
variant &
  w. avg. &
  \multicolumn{1}{c|}{w. avg.} &
  \multicolumn{1}{c|}{$1$} &
  \multicolumn{1}{c|}{$2$} &
  $3$ &
  \multicolumn{1}{c|}{w. avg.} &
  \multicolumn{1}{c|}{$1$} &
  \multicolumn{1}{c|}{$2$} &
  $3$ \\ \hline
$r\mathrm{MAE}^{model^{variant}}_{s}$ [\%] &
  \textbf{-2.9} &
  \multicolumn{1}{c|}{-0.5} &
  \multicolumn{1}{c|}{6.5} &
  \multicolumn{1}{c|}{2.1} &
  0.3 &
  \multicolumn{1}{c|}{0.5} &
  \multicolumn{1}{c|}{4.1} &
  \multicolumn{1}{c|}{5.0} &
  4.6 \\ \hline
\end{tabular}%
\label{tab:avg_results}
\end{table}
Based on the analysis performed for all considered models for lead time of 60 minutes, we will further consider only the forecasts of the cSVR model averaged according to (\ref{eq:weighted_avg}). Let us first analyse the relative errors for all deliveries $d$, all considered forecasting times $m$ and lead time of 60 minutes. The obtained results are plotted in Figure \ref{fig:3D_relative_MAE}. The cSVR model outperforms the na\"{i}ve forecast by more than $5\%$ for a~significant part of both morning and evening peaks. We observe the highest accuracy improvement for delivery starting at 17:45, where the average gain over the na\"{i}ve is $12\%$ and the maximum gain, for forecasting horizon equal to 480 minutes, is $15.5\%$. Performance similar to that in the peaks can also be observed for the first hour of the day. 
 On the other hand, the performance of the model in the middle of the day is close to that of the na\"{i}ve forecast. Lastly, performance in the late night and early morning is worse, but still close to the naive, with the overall worst relative error equal to $1.7\%$. In summary, the model outperforms the na\"{i}ve for morning and evening peaks and, less definitively, for late evening and early night hours. It does not outperform the benchmark for late night, early morning, and midday. A possible reason for this behaviour is the liquidity of the intraday market, which, as shown in Figure \ref{fig:intraday_vs_da_qtrhrly}, changes throughout the day and within hours. Comparing the bottom panel of Figure \ref{fig:intraday_vs_da_qtrhrly} with Figure \ref{fig:3D_relative_MAE}, we notice that the hours with the most accurate cSVR forecasts are those in which the number of trades with lead time over three hours is high. As a~result, more information is included in the explanatory sets, which gives an advantage over the na\"{i}ve forecast. At the same time, the periods with a~higher number of transactions with lead time over three hours are also highly volatile, see Figure \ref{fig:std_3D}. Thus, the accuracy of the na\"{i}ve forecast is further reduced. Using the proposed cSVR model in those hours allows for including more information on the market behaviour than is simply explained by the last settled price in the forecast derivation. 

\begin{figure}[!h]
    \centering
    \includegraphics[width=\textwidth]{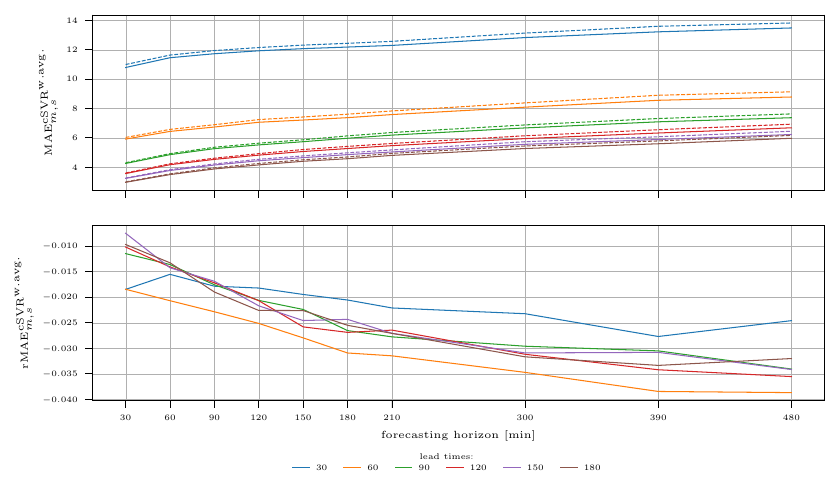}
    \caption{MAE of the cSVR model with forecast averaging (\ref{eq:weighted_avg}) (top panel, solid lines) and MAE of the na\"{i}ve forecast (top panel, dotted lines). The corresponding rMAEs are plotted in the bottom panel. The errors are averaged over all deliveries $d$.}    \label{fig:relative_and_absolute_MAE}
\end{figure}

In the next step, we extend our analysis to other lead times in the range of 30 to 180 minutes. In Figure \ref{fig:relative_and_absolute_MAE}, we show the MAE and relative MAE for the cSVR model forecasts averaged as in (\ref{eq:weighted_avg}). In the top panel, the dotted line represents the MAE of na\"{i}ve forecast, while the continuous one, for the cSVR model. Surprisingly, the highest forecast error is obtained for the shortest considered time between trade and delivery, i.e. 30 minutes. One of the possible reasons for this is the high variability of prices close to delivery, see Figure \ref{fig:std_3D}. Errors decrease with increasing lead times, most rapidly between 30 and 60 minutes. In the second panel, the relative errors for all the lead times considered are compared. The improvement in forecast accuracy when using the cSVR model is the highest for the lead time of 60 minutes and the lowest for 30 minutes lead time and forecasting horizons over 90 minutes. This means that, although the accuracy of the forecast is in general lower close to delivery due to high volatility of prices, prices of transactions settled in 60 minutes before delivery carry more information on the previous market behaviour. Recall that this is the period with the highest liquidity (see Figure \ref{fig:trades_no_3D}).    

\begin{figure}[!h]
    \centering
    \includegraphics[width=\textwidth]{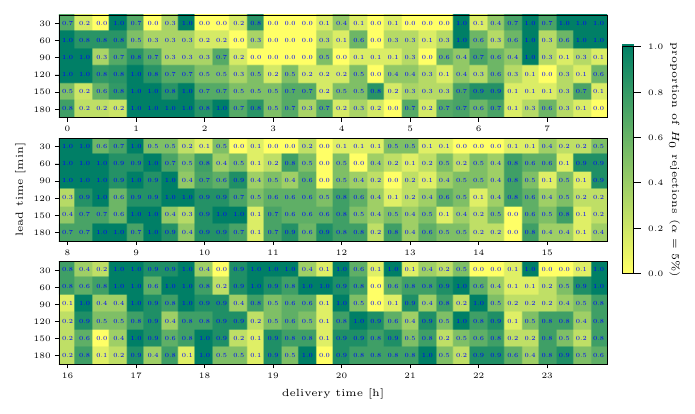}
    \caption{Proportion of the forecast horizons for which the weighted averaged cSVR model forecasts were significantly more accurate than the na\"{i}ve according the DM test (proportion of $H_0$ rejections). The significance level was set to $5\%$.}
    \label{fig:3D_DM_rejections}
\end{figure}

To further compare the accuracy of the forecast for multiple lead times, we employ the one-sided Diebold--Mariano test \cite{DM_test}. The null hypothesis is that the mean loss differential (here based on MAE) of the na\"{i}ve forecast is lower or equal to that of the cSVR model with forecast averaging. In other words, rejecting the null hypothesis means that the cSVR forecast is significantly more accurate than the na\"{i}ve. The proportions of the hypotheses rejected for the significance level of $5\%$ are plotted in Figure \ref{fig:3D_DM_rejections}. Note that in order to present the results in a~readable manner, we aggregate them over the 10 forecasting horizons considered. 
 The deliveries are split into three panels, 8 hours each. The proportion of cases in which the cSVR model is better than the na\"{i}ve for more than half of the forecasting horizons is $47\%$. The total number of cases where the cSVR model is better than the na\"{i}ve for all forecasting horizons, i.e. the value in Figure \ref{fig:3D_DM_rejections} is equal to one, is 79. This is $13.7\%$ of all 576 aggregations. Of those 79 cases, 20 and 18 occur respectively for lead times of 30 and 60 minutes. Most of the cases in which the cSVR model is significantly better than the na\"{i}ve are in the morning and evening peak hours. This corresponds to the rMAE results for 60 minutes lead time plotted in Figure \ref{fig:3D_relative_MAE}. 
 On the other hand, the total number of cases for which the number of null hypothesis rejections is $0$ is equal to $48$. From that, $23$ rejections are for a~lead time of 30 minutes and 8 and 9 for 60 and 90 minutes, respectively.

\begin{figure}[!h]
    \centering
    \includegraphics[width=\textwidth]{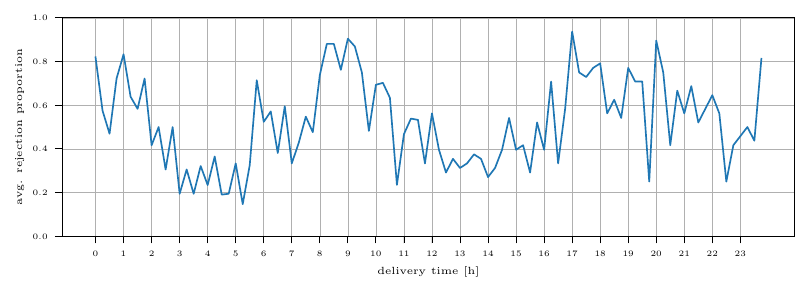}
    \caption{Proportion of the DM test null hypothesis rejections in forecasting horizons, averaged over the lead times, see also Figure \ref{fig:3D_DM_rejections} for results without averaging. 
    }
    \label{fig:aggregated_DM_rejections}
\end{figure}

Overall, the results show that a~good performance for the morning and evening peaks, early night, and late evening hours is consistent throughout the lead times. To visualise that, we aggregate the results from Figure \ref{fig:3D_DM_rejections} over lead times. The results are shown in Figure \ref{fig:aggregated_DM_rejections}. 
For deliveries between 23:45 and 1:45, and for the morning and evening peaks, the improvement is more frequent, as the model is significantly better than the na\"{i}ve for most of the forecasting horizons. 

To wrap up this section, we present a~short summary of the simulation time required by each of the considered models. Table \ref{tab:computation_time} contains the sum of the elapsed real times of computing for all forecasting horizons and chosen lead times and deliveries. Calculations were conducted on a~Windows 10 PC with 64G of 6000MHz RAM and CPU AMD Ryzen 9 7950X with a~base frequency of 4.5GHz and a~turbo clock up to 5.7GHz. The forecasting tasks, corresponding to the days in the validation window, were divided into 30 processes and run in parallel. The models were applied to each of the variables set and then averaged according to (\ref{eq:weighted_avg}). 
Clearly, the elapsed real time of computing is higher for later deliveries, as it depends greatly on the size of $\mathcal{S}^1_{d,m,s}$ variables set, increasing with the length of the price trajectories. For the same reason, the computation time is inversely proportional to the lead time. Looking at the differences between the models, RF is the most computationally demanding for all cases. LASSO is a~little faster than cSVR for short trajectories, i.e. very early deliveries, but much slower for all others. Overall, the cSVR model yielded the highest forecast accuracy while being the fastest of the models considered.
\begin{table}[!h]
\centering
\caption{Elapsed real times of computing of forecast calculations for chosen lead times and deliveries, summed over all forecasting horizons.}
\resizebox{\textwidth}{!}{%
\begin{tabular}{|l|ccc|ccc|ccc|}
\hline
lead time & \multicolumn{3}{c|}{30} & \multicolumn{3}{c|}{90} & \multicolumn{3}{c|}{180} \\ \hline
delivery  &
  \multicolumn{1}{c|}{1} &
  \multicolumn{1}{c|}{48} &
  96 &
  \multicolumn{1}{c|}{1} &
  \multicolumn{1}{c|}{48} &
  96 &
  \multicolumn{1}{c|}{1} &
  \multicolumn{1}{c|}{48} &
  96 \\ \hline
cSVR &
  \multicolumn{1}{c|}{302.85 s} &
  \multicolumn{1}{c|}{773.11 s} &
  1259.81 s &
  \multicolumn{1}{c|}{244.88 s} &
  \multicolumn{1}{c|}{738.32 s} &
  1223.45 s &
  \multicolumn{1}{c|}{185.91 s} &
  \multicolumn{1}{c|}{694.65 s} &
  1055.13 s \\ \hline
LASSO &
  \multicolumn{1}{c|}{204.88 s} &
  \multicolumn{1}{c|}{2595.73 s} &
  3788.57 s &
  \multicolumn{1}{c|}{154.13 s} &
  \multicolumn{1}{c|}{2392.96 s} &
  4669.91 s &
  \multicolumn{1}{c|}{103.30 s} &
  \multicolumn{1}{c|}{1915.77 s} &
  4317.45 s \\ \hline
RF &
  \multicolumn{1}{c|}{1370.84 s} &
  \multicolumn{1}{c|}{4720.88 s} &
  5969.73 s &
  \multicolumn{1}{c|}{985.44 s} &
  \multicolumn{1}{c|}{4146.05 s} &
  7107.30 s &
  \multicolumn{1}{c|}{667.20 s} &
  \multicolumn{1}{c|}{3617.79 s} &
  8841.99 s \\ \hline
\end{tabular}%
}
\label{tab:computation_time}
\end{table}

\section{Summary and discussion}
\label{sec:summary}
In this paper, we have developed and tested a~kernel method for point forecasting in the intraday continuous power market. The approach is based on a~kernel function correction inspired by the recent Neural Tangent Kernels (NTKs) developments and aims at introducing the local elasticity idea to kernel methods by making the kernels directly dependent on the response variable. 
While a~similar kernel modification was previously used in classification studies, it is, to our best knowledge, the first direct implementation of this idea in time series forecasting scheme. 
The SVR model with kernel correction based on the na\"{i}ve forecast, denoted as cSVR, was tested for the German continuous market data.
We have shown that the cSVR model performs better than the three considered benchmarks: LASSO, RF, and the na\"{i}ve forecast. We conclude that the cSVR, being a~nonparametric, distribution-free method, is more robust and suitable to deal with continuous market price trajectories, which characteristic changes drastically throughout the deliveries and forecasting time. 

We trained the proposed method on three sets of expert explanatory variables. The first set is based on the exogenous fundamental variables and a~general information from the transactions of the delivery of interest and adjacent deliveries. The second set is constructed using only the information from the last 60 minutes on the delivery of interest and the last four deliveries. Lastly, the third set of variables contains the exogenous variables plus only the most recent price information. The best results for individual expert sets were obtained for the third one. Thus, we conclude that the exogenous variables carry important information about prices even in a~very short-term forecasting scheme. We have shown that the average taken from the forecasts of three cSVR models, trained separately on three sets of explanatory variables, and the na\"{i}ve, improves the accuracy over each individual forecast. Moreover, we used the weighting scheme for the forecasts averaging and tested it. Although the results of the weighted averaging were not significantly better than those given by a~simple arithmetic average, we suggest using the weighting scheme, as it is not computationally expensive and adds robustness to the model.

The forecasting accuracy of the averaged cSVR model was tested for several lead times, forecasting horizons, and all deliveries. We have shown that the accuracy is highest for deliveries in the morning and evening peaks, for which the cSVR model outperforms the na\"{i}ve even by $15.5\%$. For all deliveries in the morning, that is, from 7:00 to 10:45, the average gain over the na\"{i}ve, from all forecast horizons, is $3\%$. Similarly, for evening deliveries, from 19:00 to 23:45, the cSVR model outperforms the na\"{i}ve by $3.3\%$. We argue that the good performance in the peaks is caused by the higher liquidity for larger lead times in those hours. The forecasting study for the German intraday continuous market dataset also showed that the forecasting error is inversely proportional to the lead time, i.e., the time between trade and delivery. This is caused by the increasing volatility of prices closer to the delivery, but also by the data limitations. As we do not have access to the order book, the price for more distant lead times is often inferred based on the most recent transaction. Thus, it can be constant for several minutes or even hours. This further dampens volatility and causes the error to decrease. The impact of this limitation would be significantly greater for lead times of more than 180 minutes.

Our approach can be further studied and expanded. From an application perspective, the data used in the forecasting study conducted in this paper are only a~small part of market information. Access to orders or even their volume-weighted averages would greatly enhance the study, allowing one to build more detailed price trajectories and perform an in-depth analysis of liquidity. Another practical remark is that we did not use any information on half-hourly and hourly products. By incorporating them into the model, one could achieve better accuracy by extracting more information from the market. As for the method itself, by kernel correction based on the multiplication with na\"{i}ve based kernel, only the damping of outlier points is performed. Although it is in line with the broader understanding of local elasticity, it would also be beneficial to reduce the influence of the leverage points in the kernel. We already noticed that the Laplace kernel using the $L2$ norm performs better than the $L1$ norm Laplace kernel. This suggests that a~more rapid damping of the kernel values is plausible in case of continuous markets. The development of a~tailored kernel that manages to better catch dissimilarities would be a~step to further increase the accuracy of our model. Another important issue is the hand-picking of several constants throughout the study. We acknowledge that cross-validation should be performed for RF hyperparameters, SVR model hyperparameters, and optimally even the kernel widths. However, due to the computational burden, we resorted to the hand-picked values. Developing methods to derive the hyperparameters of SVR, based on data characteristics, would be a~step towards enhancing the accuracy of the model. Our contribution in this direction is the calculation of the kernel widths based on the quantiles of the distances matrix. 

We believe that a~method outperforming well-established benchmarks would be usable in practice to obtain trading signals in the intraday continuous electricity markets.

\section*{Acknowledgements}
\noindent The authors acknowledge the Wrocław Centre for Networking and Supercomputing (WCSS) for providing computational resources that supported this work.

\vspace{11pt}
\noindent \textbf{Author contributions:}

\noindent Andrzej Puć: conceptualisation, formal analysis, methodology, software, writing - original draft,

\noindent Joanna Janczura: supervision, conceptualisation, resources, writing - review \& editing.

\bibliographystyle{abbrv}
\bibliography{bibliography} 

\begin{thebibliography}{10}

\bibitem{sara_atef__2019}
S.~Atef and A.~B. Eltawil.
\newblock A comparative study using deep learning and support vector regression
  for electricity price forecasting in smart grids.
\newblock In {\em 2019 IEEE 6th International Conference on Industrial
  Engineering and Applications (ICIEA)}, pages 603--607, 2019.

\bibitem{bai2020taylorized}
Y.~Bai, B.~Krause, H.~Wang, C.~Xiong, and R.~Socher.
\newblock Taylorized training: Towards better approximation of neural network
  training at finite width, 2020.

\bibitem{random_forest}
L.~Breiman.
\newblock Random forests.
\newblock {\em Machine Learning}, 45(1):5–32, 2001.

\bibitem{svr_equation}
C.-C. Chang and C.-J. Lin.
\newblock Libsvm: A library for support vector machines.
\newblock {\em ACM Trans. Intell. Syst. Technol.}, 2(3), 2011.

\bibitem{label_aware_ntk}
S.~Chen, H.~He, and W.~J. Su.
\newblock Label-aware neural tangent kernel: Toward better generalization and
  local elasticity.
\newblock {\em CoRR}, abs/2010.11775, 2020.

\bibitem{DM_test}
F.~Diebold and R.~Mariano.
\newblock Comparing predictive accuracy.
\newblock {\em Journal of Business and Economic Statistics}, 13(3):253--63,
  1995.

\bibitem{NIPS1996_d3890178}
H.~Drucker, C.~J.~C. Burges, L.~Kaufman, A.~Smola, and V.~Vapnik.
\newblock Support vector regression machines.
\newblock In M.~Mozer, M.~Jordan, and T.~Petsche, editors, {\em Advances in
  Neural Information Processing Systems}, volume~9. MIT Press, 1996.

\bibitem{efron_lars}
B.~Efron, T.~Hastie, I.~Johnstone, and R.~Tibshirani.
\newblock Least angle regression.
\newblock {\em The Annals of Statistics}, 32(2):407--451, 2004.

\bibitem{ENTSOE}
European association for the cooperation of transmission system operators
  (tsos) for electricity.
\newblock \url{https://transparency.entsoe.eu/}.
\newblock Accessed: 2021-04-12.

\bibitem{physical_flows_data_avails}
ENTSOE.
\newblock Physical flows [12.1.g].
\newblock
  \url{https://transparency.entsoe.eu/content/static_content/Static%20content/knowledge%20base/data-views/transmission-domain/Data-view%20Cross%20Border%20Physical%20Flows.html}.
\newblock Accessed: 2024-03-17.

\bibitem{epex_transactions_delay}
EPEX.
\newblock Epex spot sftp server file specifications.
\newblock
  \url{https://www.epexspot.com/sites/default/files/download_center_files/SFTP_specifications_2020-07.pdf}.
\newblock Accessed: 2024-10-17.

\bibitem{EPEX}
{European Energy Exchange}.
\newblock \url{https://www.epexspot.com}.
\newblock Accessed: 2021-10-25.

\bibitem{EPEX_market_design}
Basics of the power market.
\newblock \url{https://www.epexspot.com/en/basicspowermarket}.
\newblock Accessed: 2024-10-21.

\bibitem{NTK_vs_DNN}
S.~Fort, G.~K. Dziugaite, M.~Paul, S.~Kharaghani, D.~M. Roy, and S.~Ganguli.
\newblock Deep learning versus kernel learning: an empirical study of loss
  landscape geometry and the time evolution of the neural tangent kernel.
\newblock {\em CoRR}, abs/2010.15110, 2020.

\bibitem{laplace_NTK_similarity}
A.~Geifman, A.~K. Yadav, Y.~Kasten, M.~Galun, D.~W. Jacobs, and R.~Basri.
\newblock On the similarity between the {Laplace} and neural tangent kernels.
\newblock {\em CoRR}, abs/2007.01580, 2020.

\bibitem{GIANFREDA20122228}
A.~Gianfreda and L.~Grossi.
\newblock Forecasting {Italian} electricity zonal prices with exogenous
  variables.
\newblock {\em Energy Economics}, 34(6):2228--2239, 2012.

\bibitem{AG}
A.~Gianfreda, L.~Parisio, and M.~Pelagatti.
\newblock The impact of {RES} in the {I}talian day-ahead and balancing markets.
\newblock {\em Energy Journal}, 37:161--184, 2016.

\bibitem{germany_exchange}
M.~C. group.
\newblock Electricity in germany.
\newblock
  \url{https://oec.world/en/profile/bilateral-product/electricity/reporter/deu?redirect=true&yearExportSelector=exportYear3}.
\newblock Accessed: 2024-03-17.

\bibitem{Local_elasticity}
H.~He and W.~J. Su.
\newblock The local elasticity of neural networks.
\newblock {\em CoRR}, abs/1910.06943, 2019.

\bibitem{NN_in_point_time_series}
H.~Hewamalage, C.~Bergmeir, and K.~Bandara.
\newblock Recurrent neural networks for time series forecasting: Current status
  and future directions.
\newblock {\em International Journal of Forecasting}, 37(1):388--427, 2021.

\bibitem{hirsch2023multivariatesimulationbasedforecastingintraday}
S.~Hirsch and F.~Ziel.
\newblock Multivariate simulation-based forecasting for intraday power markets:
  Modeling cross-product price effects.
\newblock {\em Applied Stochastic Models in Business and Industry}, n/a(n/a),
  2024.

\bibitem{Hirsh_simulation_based}
S.~Hirsch and F.~Ziel.
\newblock Simulation-based forecasting for intraday power markets: Modelling
  fundamental drivers for location, shape and scale of the price distribution.
\newblock {\em The Energy Journal}, 45(3):87--124, 2024.

\bibitem{hong_weron_review_2020}
T.~Hong, P.~Pinson, Y.~Wang, R.~Weron, D.~Yang, and H.~Zareipour.
\newblock Energy forecasting: A review and outlook.
\newblock {\em IEEE Open Access Journal of Power and Energy}, 7:376--388, 2020.

\bibitem{Hubicka2019ANO}
K.~Hubicka, G.~Marcjasz, and R.~Weron.
\newblock A note on averaging day-ahead electricity price forecasts across
  calibration windows.
\newblock {\em IEEE Transactions on Sustainable Energy}, 10:321--323, 2019.

\bibitem{kernel_width_based_on_median}
T.~Jaakkola, M.~Diekhans, and D.~Haussler.
\newblock Using the {Fisher} kernel method to detect remote protein homologies.
\newblock In {\em Proceedings of the Seventh International Conference on
  Intelligent Systems for Molecular Biology}, page 149–158. AAAI Press, 1999.

\bibitem{jacot_ntk}
A.~Jacot, F.~Gabriel, and C.~Hongler.
\newblock Neural tangent kernel: Convergence and generalization in neural
  networks.
\newblock {\em CoRR}, abs/1806.07572, 2018.

\bibitem{janczura_puc_strategie}
J.~Janczura and A.~Puć.
\newblock {ARX-GARCH} probabilistic price forecasts for diversification of
  trade in electricity markets—variance stabilizing transformation and
  financial risk-minimizing portfolio allocation.
\newblock {\em Energies}, 16(2), 2023.

\bibitem{Kath_2018}
C.~Kath and F.~Ziel.
\newblock The value of forecasts: Quantifying the economic gains of accurate
  quarter-hourly electricity price forecasts.
\newblock {\em Energy Economics}, 76:411–423, 2018.

\bibitem{KIESEL201777}
R.~Kiesel and F.~Paraschiv.
\newblock Econometric analysis of 15-minute intraday electricity prices.
\newblock {\em Energy Economics}, 64:77--90, 2017.

\bibitem{Kulakov_Ziel}
S.~Kulakov and F.~Ziel.
\newblock The impact of renewable energy forecasts on intraday electricity
  prices.
\newblock {\em Economics of Energy $\&$ Environmental Policy}, 10, 2021.

\bibitem{Weron_review}
J.~Lago, G.~Marcjasz, B.~{De Schutter}, and R.~Weron.
\newblock Forecasting day-ahead electricity prices: A review of
  state-of-the-art algorithms, best practices and an open-access benchmark.
\newblock {\em Applied Energy}, 293:116983, 2021.

\bibitem{NN_vs_SVR}
S.~Luo and Y.~Weng.
\newblock {A two-stage supervised learning approach for electricity price
  forecasting by leveraging different data sources}.
\newblock {\em Applied Energy}, 242(C):1497--1512, 2019.

\bibitem{Maciejowska_EE}
K.~Maciejowska.
\newblock {Assessing the impact of renewable energy sources on the electricity
  price level and variability – A quantile regression approach}.
\newblock {\em Energy Economics}, 85:104532, 2020.

\bibitem{maciejo_PCA_averaging}
K.~Maciejowska, B.~Uniejewski, and T.~Serafin.
\newblock {PCA} forecast averaging—predicting day-ahead and intraday
  electricity prices.
\newblock {\em Energies}, 13(14), 2020.

\bibitem{marcjasz_NN_probabilistic}
G.~Marcjasz, M.~Narajewski, R.~Weron, and F.~Ziel.
\newblock Distributional neural networks for electricity price forecasting.
\newblock {\em Energy Economics}, 125:106843, 2023.

\bibitem{averaging_forecasts}
G.~Marcjasz, T.~Serafin, and R.~Weron.
\newblock Selection of calibration windows for day-ahead electricity price
  forecasting.
\newblock {\em Energies}, 11(9), 2018.

\bibitem{beating_the_naive}
G.~Marcjasz, B.~Uniejewski, and R.~Weron.
\newblock Beating the naïve—combining {LASSO} with naïve intraday
  electricity price forecasts.
\newblock {\em Energies}, 13(7), 2020.

\bibitem{AM}
C.~B. Martinez-Anido, G.~Brinkman, and B.~Hodge.
\newblock The impact of wind power on electricity prices.
\newblock {\em Renewable Energy}, 94:474--487, 2016.

\bibitem{NARAJEWSKI2020100107}
M.~Narajewski and F.~Ziel.
\newblock Econometric modelling and forecasting of intraday electricity prices.
\newblock {\em Journal of Commodity Markets}, 19:100107, 2020.

\bibitem{NARAJEWSKI2020115801}
M.~Narajewski and F.~Ziel.
\newblock Ensemble forecasting for intraday electricity prices: Simulating
  trajectories.
\newblock {\em Applied Energy}, 279:115801, 2020.

\bibitem{Next-Kraftwerke}
Intraday trading: Definition, theory and practice.
\newblock \url{https://www.next-kraftwerke.com/knowledge/intraday-trading}.
\newblock Accessed: 2024-08-11.

\bibitem{forecasts_combination_weron_nowotarski}
J.~Nowotarski and R.~Weron.
\newblock {To combine or not to combine? Recent trends in electricity price
  forecasting}.
\newblock HSC Research Reports HSC/16/01, Hugo Steinhaus Center, Wroclaw
  University of Technology, 2016.

\bibitem{scikit-learn}
F.~Pedregosa, G.~Varoquaux, A.~Gramfort, V.~Michel, B.~Thirion, O.~Grisel,
  M.~Blondel, P.~Prettenhofer, R.~Weiss, V.~Dubourg, J.~Vanderplas, A.~Passos,
  D.~Cournapeau, M.~Brucher, M.~Perrot, and E.~Duchesnay.
\newblock Scikit-learn: Machine learning in {P}ython.
\newblock {\em Journal of Machine Learning Research}, 12:2825--2830, 2011.

\bibitem{representer_theorem}
B.~Sch{\"o}lkopf, R.~Herbrich, and A.~J. Smola.
\newblock A generalized representer theorem.
\newblock In D.~Helmbold and B.~Williamson, editors, {\em Computational
  Learning Theory}, pages 416--426, Berlin, Heidelberg, 2001. Springer Berlin
  Heidelberg.

\bibitem{scikit_LASSOLARS}
{scikit-learn}.
\newblock Lassolarscv.
\newblock
  \url{https://scikit-learn.org/stable/modules/generated/sklearn.linear_model.LassoLarsCV.html}.
\newblock Accessed: 2024-08-13.

\bibitem{scikit_RF}
{scikit-learn}.
\newblock Randomforestregressor.
\newblock
  \url{https://scikit-learn.org/stable/modules/generated/sklearn.ensemble.RandomForestRegressor.html}.
\newblock Accessed: 2024-08-13.

\bibitem{scikit_SVR}
{scikit-learn}.
\newblock Svr.
\newblock
  \url{https://scikit-learn.org/stable/modules/generated/sklearn.svm.SVR.html}.
\newblock Accessed: 2024-06-16.

\bibitem{SERAFIN2022106125}
T.~Serafin, G.~Marcjasz, and R.~Weron.
\newblock Trading on short-term path forecasts of intraday electricity prices.
\newblock {\em Energy Economics}, 112:106125, 2022.

\bibitem{SIDC}
Single intraday coupling (sidc).
\newblock \url{https://www.entsoe.eu/network_codes/cacm/implementation/sidc/}.
\newblock Accessed: 2024-10-21.

\bibitem{lasso}
R.~Tibshirani.
\newblock Regression shrinkage and selection via the lasso: a retrospective.
\newblock {\em Journal of the Royal Statistical Society. Series B (Statistical
  Methodology)}, 73(3):273--282, 2011.

\bibitem{TSCHORA2022118752}
L.~Tschora, E.~Pierre, M.~Plantevit, and C.~Robardet.
\newblock Electricity price forecasting on the day-ahead market using machine
  learning.
\newblock {\em Applied Energy}, 313:118752, 2022.

\bibitem{UNIEJEWSKI2019}
B.~Uniejewski, G.~Marcjasz, and R.~Weron.
\newblock Understanding intraday electricity markets: Variable selection and
  very short-term price forecasting using {LASSO}.
\newblock {\em International Journal of Forecasting}, 35(4):1533--1547, 2019.

\end{thebibliography}

\end{document}